\RequirePackage{lineno}
\documentclass[aps,prl,twocolumn,superscriptaddress,floatfix]{revtex4-1}

\usepackage{graphicx}
\usepackage{amsmath,amssymb}
\usepackage{hyperref}
\usepackage{datetime}
\usepackage[utf8]{inputenc}
\usepackage{braket}
\usepackage{floatrow}
\usepackage{comment}

\begin{document}

\title{Probing entanglement in a many-body-localized system}

\author{Alexander~Lukin}
\affiliation{Department of Physics, Harvard University, Cambridge, Massachusetts 02138, USA}
\author{Matthew~Rispoli}
\affiliation{Department of Physics, Harvard University, Cambridge, Massachusetts 02138, USA}
\author{Robert~Schittko}
\affiliation{Department of Physics, Harvard University, Cambridge, Massachusetts 02138, USA}
\author{ M.~Eric~Tai}
\affiliation{Department of Physics, Harvard University, Cambridge, Massachusetts 02138, USA}
\author{Adam~M.~Kaufman}
\affiliation{Department of Physics, Harvard University, Cambridge, Massachusetts 02138, USA}
\affiliation{JILA, National Institute of Standards and Technology and University of Colorado, and Department of Physics, University of Colorado, Boulder, Colorado 80309, USA}
\author{Soonwon~Choi}
\affiliation{Department of Physics, Harvard University, Cambridge, Massachusetts 02138, USA}
\author{Vedika~Khemani}
\affiliation{Department of Physics, Harvard University, Cambridge, Massachusetts 02138, USA}
\author{Julian~L\'eonard}
\affiliation{Department of Physics, Harvard University, Cambridge, Massachusetts 02138, USA}
\author{Markus~Greiner}
\email{greiner@physics.harvard.edu}
\affiliation{Department of Physics, Harvard University, Cambridge, Massachusetts 02138, USA}

\begin{abstract}
An interacting quantum system that is subject to disorder may cease to thermalize due to localization of its constituents, thereby marking the breakdown of thermodynamics. The key to our understanding of this phenomenon lies in the system's entanglement, which is experimentally challenging to measure. We realize such a many-body-localized system in a disordered Bose-Hubbard chain and characterize its entanglement properties through particle fluctuations and correlations. We observe that the particles become localized, suppressing transport and preventing the thermalization of subsystems. Notably, we measure the development of non-local correlations, whose evolution is consistent with a logarithmic growth of entanglement entropy - the hallmark of many-body localization. Our work experimentally establishes many-body localization as a qualitatively distinct phenomenon from localization in non-interacting, disordered systems.
\end{abstract}

\maketitle

\date{\today}
\section*{Introduction}

Isolated quantum many-body systems, undergoing unitary time evolution, maintain their initial global purity. However, the presence of interactions drives local thermalization: the coupling between any subsystem and its remainder mimics the contact with a bath. This causes the subsystem's degrees of freedom to be ultimately described by a thermal ensemble, even if the full system is in a pure state \cite{Deutsch1991, Srednicki1994, Rigol2008, Neill2016, Kaufman2016}. A consequence of thermalization is that local information about the initial state of the subsystem gets scrambled and transferred into non-local correlations that are only accessible through global observables \cite{Neill2016, Kaufman2016, Nandkishore2015}.

\begin{figure}[h!]
	\includegraphics[scale= 0.94]{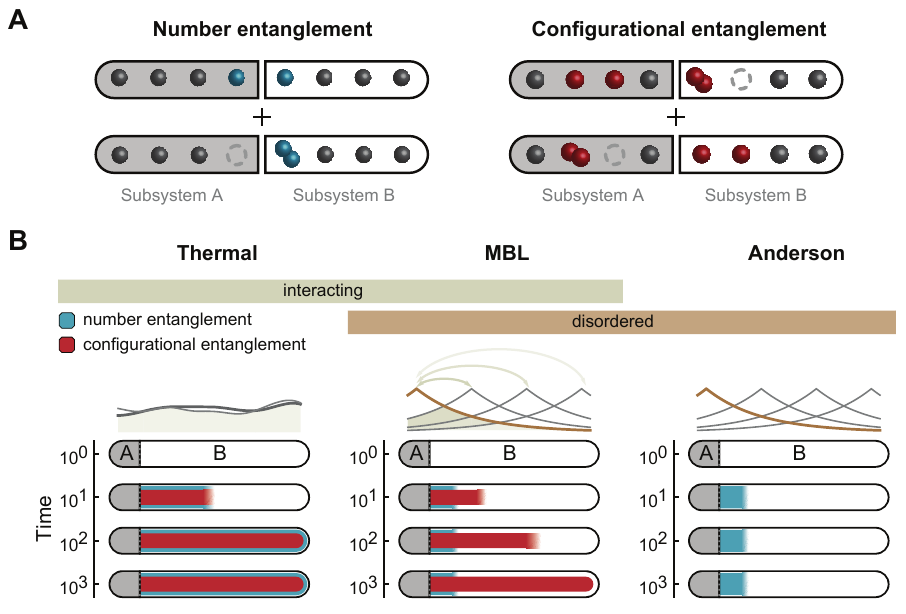}
	\caption{{\bf Entanglement dynamics in non-equilibrium quantum systems. (A)} Subsystems A and B of an isolated system out of equilibrium entangle in two different ways: \textit{number entanglement} stems from a superposition of states with different particle numbers in the subsystems and is generated through particle motion across the boundary; \textit{configurational entanglement} stems from a superposition of states with different particle arrangement within the subsystems and requires both particle motion and interactions. \textbf{(B)} In the absence of disorder, both types of entanglement rapidly spread across the entire system due to delocalization of particles (left panel). The degree of entanglement and the timescales change drastically when applying disorder (central panel): particle localization spatially restricts number entanglement, yet interactions allow configurational entanglement to form very slowly across the entire system. A disordered system without interactions shows only local number entanglement while the slow growth of configurational entanglement is completely absent (right panel).}
	\label{fig:schematics}
\end{figure}

Disordered systems \cite{Anderson1958, Wiersma1997,  Schwartz2007, Billy2008, Roati2008, Lahini2008, Deissler2010, Gadway2011, DErrico2014, Kondov2015} can provide an exception to this paradigm of quantum thermalization. In such systems, particles can localize and transport ceases, which prevents thermalization. This phenomenon is called many-body localization (MBL) \cite{Nandkishore2015, Anderson1958, Gornyi2005, Basko2006, Oganesyan2007, Pal2010, Imbrie2016, Abanin2018}. Experimental studies have identified MBL through the persistence of the initial density distribution \cite{Abanin2018, Schreiber2015, Smith2015, Choi2016} and two-point correlation functions during transient dynamics \cite{Smith2015}. However, while particle transport is frozen, the presence of interactions gives rise to slow coherent many-body dynamics that generate non-local correlations, which are inaccessible to local observables \cite{Serbyn2013, Serbyn2013a, Huse2014}. These dynamics are considered to be the hallmark of MBL and distinguish it from its non-interacting counterpart, called Anderson localization \cite{Anderson1958, Wiersma1997, Schwartz2007, Billy2008, Roati2008, Lahini2008}. Their observation, however, has remained elusive.

\begin{figure*}[t]
	\includegraphics[scale= 0.94]{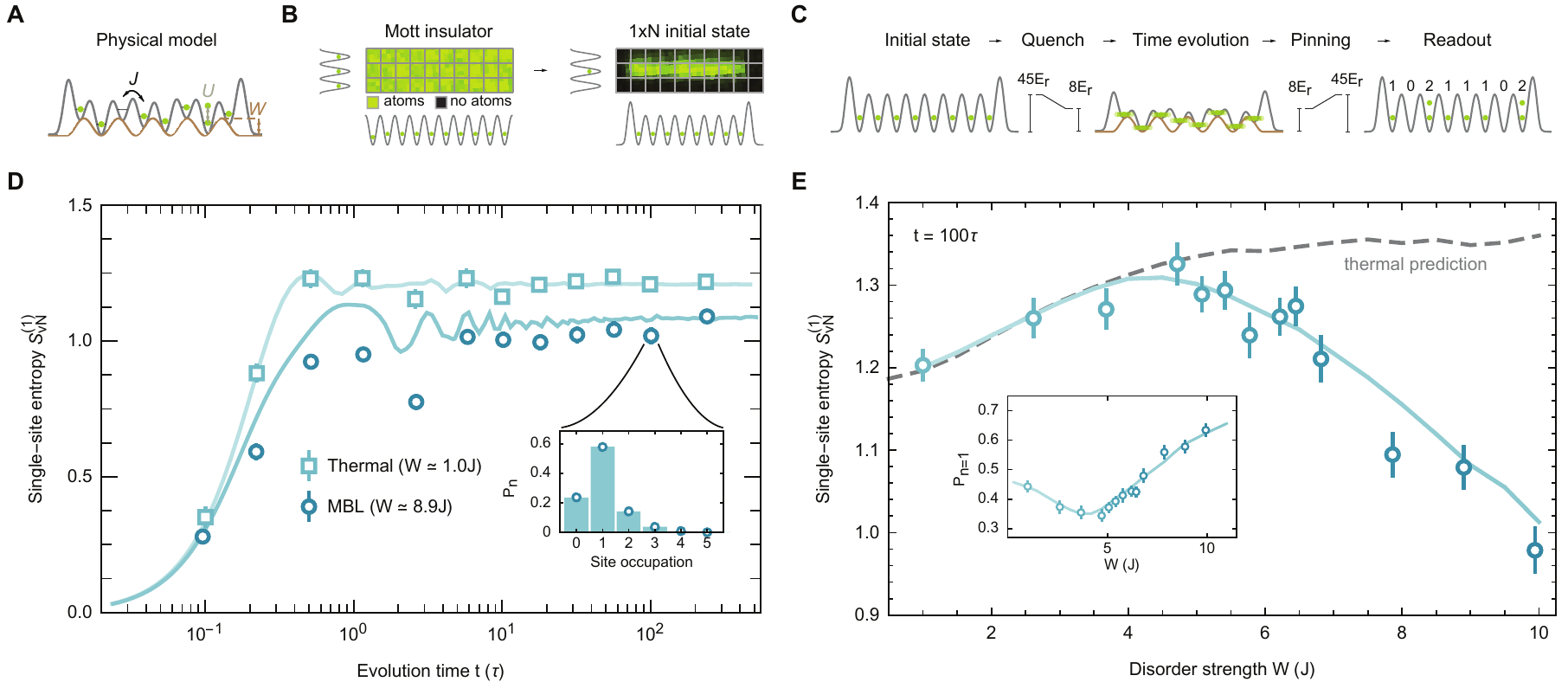}
	\caption{{\bf Site-resolved measurement of thermalization breakdown. (A)} One-dimensional Aubry-Andr\'e model with particle tunneling at rate $J/\hbar$, on-site interaction energy $U$ and quasi-periodic potential with amplitude $W$. {\bf(B)} We prepare the initial state of eight unentangled atoms by projecting tailored optical potentials on a two-dimensional Mott insulator at $45E_\text{r}$ lattice depth, where $E_\text{r}/h = 1.24\,\text{kHz}$ is the recoil energy. {\bf (C)} We create a non-equilibrium system by abruptly enabling tunneling dynamics. Following a variable evolution time, we project the many-body state back onto the number basis by increasing the lattice depth, and obtain the site-resolved atom number from a fluorescence image (SI). {\bf(D)} We compute the single-site von Neumann entropy $S_\text{vN}^{(1)}$ from the site-resolved atom number statistics (inset) after different evolution times (scaled with tunneling time $\tau=\hbar/J$) in the presence of weak and strong disorder. {\bf(E)} Probability $p_1$ to retrieve the initial state (inset) and $S_\text{vN}^{(1)}$ for different $W$, measured after $100\tau$ evolution. The deviation from the thermal ensemble prediction for strong disorder signals the breakdown of thermalization in the system. All lines in {\bf (C-D)} show the prediction of exact diagonalization calculations without any free parameters. Each data point is sampled from 197 disorder realizations (SI). Error bars denote the s.e.m.}
	\label{fig:dynamics}
\end{figure*}

We study these many-body dynamics by probing the entanglement properties of an MBL system with fixed particle number \cite{Serbyn2013, Serbyn2013a, Huse2014, Znidaric2008, Bardarson2012}. We distinguish two types of entanglement that can exist between a subsystem and its complement (Fig.~1A): \textit{Number entanglement} implies that the particle number in one subsystem is correlated with the particle number in the other. It is generated through tunneling across the boundary between the subsystems. \textit{Configurational entanglement} implies that the configuration of the particles in one subsystem is correlated with the configuration of the particles in the other. It arises from a combination of particle motion and interaction. The formation of particle and configurational entanglement changes in the presence or absence of interactions and disorder in the system (Fig.~1B). In thermal systems without disorder, interacting particles delocalize and rapidly create both types of entanglement throughout the entire system. Contrarily, for Anderson localization, number entanglement builds up only locally at the boundary between the two subsystems. Here the lack of interactions prevents the substantial formation of configurational entanglement. In MBL systems, number entanglement builds up in a similarly local way as for Anderson localization. However, notably, the presence of interactions additionally enables the slow formation of configurational entanglement throughout the entire system. 

In this work, we realize an MBL system and characterize these key properties: breakdown of quantum thermalization, finite localization length of the particles, area-law scaling of the number entanglement, and slow growth of the configurational entanglement that ultimately results in a volume-law scaling. Each property shows a contrasting behavior when the system is prepared at weak disorder in a thermalizing state. While the former three properties are also present for an Anderson localized state, the slowly growing configurational entanglement qualitatively distinguishes our system from a non-interacting, localized state. 

\section*{Experimental system}

In our experiments, we study MBL in the interacting Aubry-Andr\'e model for bosons in one dimension \cite{Aubry1980, Iyer2013}, which is described by the Hamiltonian
\begin{equation}
\begin{split}
\hat{\mathcal{H}} =  &-J \sum_i \left(\hat{a}_i^\dagger\hat{a}_{i+1} + h.c.\right) \\
&+ \frac{U}{2} \sum_i\hat{n}_i \left( \hat{n}_i - 1\right) + W\sum_i h_i\hat{n}_i\text{,}
\end{split}
\label{eq:H_AA}
\end{equation}

where $\hat{a}_i^\dagger$ ($\hat{a}_i$) is the creation (annihilation) operator for a boson on site $i$, and $\hat{n}_i = \hat{a}_i^\dagger\hat{a}_i$ is the particle number operator on that site. The first term describes the tunneling between neighboring lattice sites with the rate $J/\hbar$, where $\hbar$ is the reduced Planck constant. The second term represents the energy shift $U$ when multiple particles occupy the same site. The last term introduces a site-resolved potential offset, which is created with an incommensurate lattice $h_i = \cos\left(2\pi\beta i+\phi\right)$ of period $\beta\approx 1.618$ lattice sites, phase $\phi$, and amplitude $W$. In our experiment, we achieve independent control over $J$, $W$, and $\phi$ (Fig.~2A).

Our experiments begin with a Mott-insulating state in the atomic limit with one $^{87}$Rb atom on each site of a two-dimensional optical lattice (Fig.~2B). The system is placed in the focus of a high-resolution imaging system through which we project site-resolved optical potentials \cite{Bakr2009}. We first isolate a single, one-dimensional chain from the Mott insulator and then add the site-resolved potential offsets $W_i$ with the incommensurate lattice. At this point, the system remains in a product state of one atom per lattice site. We abruptly switch on the tunneling by reducing the lattice depth within a fraction of the tunneling time (Fig.~2C). This quench brings the system to a non-equilibrium state and initializes the unitary time dynamics corresponding to the above Hamiltonian. The tunneling time $\tau =\hbar/J = 4.3(1)\,\text{ms}$ and the interaction strength $U = 2.87(3) J$ remain constant in all our experiments. Following a variable evolution time, we abruptly increase the lattice depth to project the many-body state back onto the number basis, which consists of all possible distributions of the particles within the chain. Finally, we image the system in an atom-number-sensitive way with single-site resolution (SI). 

In some realizations, particle loss during the time evolution and imperfect readout reduce the number of detected atoms compared to the initial state, thereby injecting classical entropy into the system. We eliminate this entropy by post-selecting the data on the intended atom number, thereby reaching a fidelity of $99.1(2)\%$ unity filling in the initial state, which is limited by the fraction of doublon-hole pairs in the Mott insulator. The result is a highly pure state, in which all correlations are expected to stem from entanglement in the system. 

\section*{Breakdown of thermalization}

We first investigate the breakdown of thermalization in a subsystem that consists of a single lattice site. The conserved total atom number enforces a one-to-one correspondence between the particle number outcome on a single site and the number in the remainder of the system---entangling the two during tunneling dynamics. Ignoring information about the remaining system puts the subsystem into a mixed state of different number states. The associated number entropy is given by $S_\text{n}^{(1)} = -\sum_{n} p_n \log(p_n)$, where $p_n$ is the probability of finding $n$ atoms in the subsystem (SI). Since the atom number is the only degree of freedom of a single lattice site, $S_\text{n}^{(1)}$ captures all of the entanglement between the subsystem and its complement, and is equivalent to the single-site von Neumann entanglement entropy $S_\text{vN}^{(1)}$.

Counting the atom number on an individual lattice site in different experimental realizations allows us to obtain the probabilities $p_n$ and compute $S_\text{vN}^{(1)}$. We perform such measurements for various evolution times. At low disorder depth ($W=1.0(1)J$), the entropy grows over a few tunneling times and then reaches a stationary value (Fig.~2D). The stationary value is reduced for deep disorder ($W=8.9(1)J$) and remains constant over two orders of magnitude, up to several hundred tunneling times. The lack of entropy increase indicates the absence of heating in the system. The excellent agreement of the measured entropy with \textit{ab initio} calculations up to the longest measured evolution times suggests a highly unitary evolution of the system.

We perform measurements of $S_\text{vN}^{(1)}$ at different disorder strengths following an evolution of one hundred tunneling times (Fig.~2E). To evaluate the degree of local thermalization, we compare the results with 
the prediction of a thermal ensemble for our system (SI). For weak disorder, the measured entropy agrees with the predicted value, whereas the entropy is significantly reduced for strong disorder---signaling the absence of thermalization in the system. As a consequence, the system retains some memory of its initial conditions for arbitrarily long evolution times. We indeed find that the probability to retrieve the initial state of one atom per site increases for strong disorder (inset Fig.~2E).

\begin{figure}
	\centering
	\includegraphics[scale= 0.94]{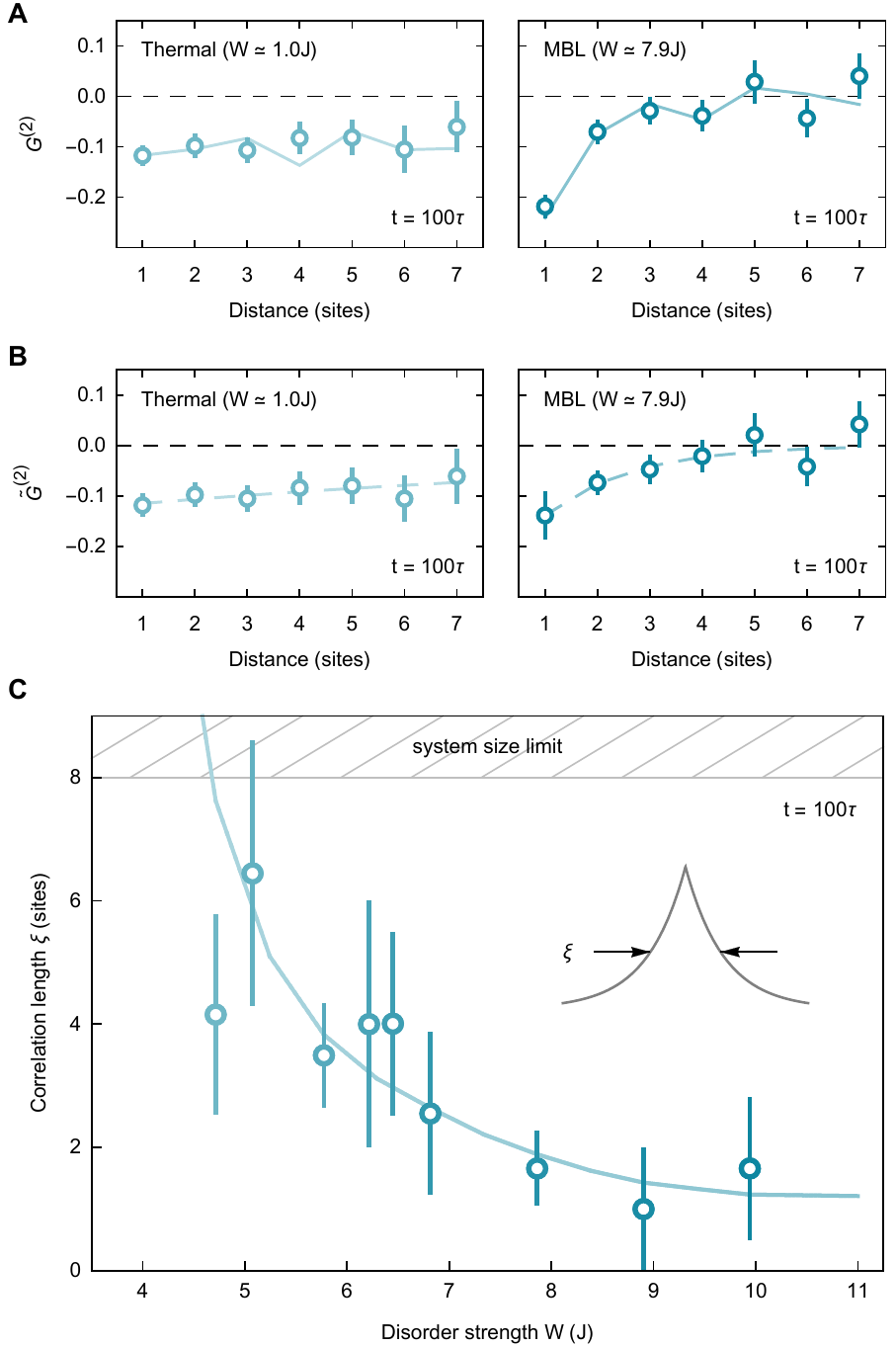}
	\caption{{\bf Spatial localization of particles.} {\bf{(A)}} The density-density correlations $G^{(2)} (d)$ as a function of distance $d$ at weak and strong disorder after an evolution time of $100\tau$. The alternating nature of the density-density correlations stems from the autocorrelation function of the quasiperiodic potential. {\bf (B)} Subtracting the influence of the quasiperiodic potential (SI) reveals the exponential decay of the correlation function. {\bf (C)} Particle motion is confined within the correlation length $\xi$. We use a fit to extract $\xi$ for different disorder strengths. The fit function is a product of an exponential decay with the autocorrelation function of the quasiperiodic potential (SI). Each measurement is sampled from 197 disorder realizations (SI). The solid lines show the prediction of exact diagonalization---calculated without any free parameters. Error bars denote the s.e.m in {\bf (A-B)}, and the fit error in {\bf (C)}.}
	\label{fig:localization}
\end{figure}

\section*{Spatial localization}

The breakdown of thermalization is expected to be a consequence of the spatial localization of the particles. Previous experiments have determined the decay length of an initially prepared density step into empty space \cite{Choi2016}. We measure the localization by directly probing density-density correlations within the system. These correlations are captured by $G^{(2)}(d)=\langle n_{i} n_{i+d} \rangle - \langle n_{i}\rangle \langle n_{i+d}\rangle$, where $\langle...\rangle$ denotes averaging over different disorder realizations as well as all sites $i$ of the chain. The particle numbers on two sites at distance $d>0$ are uncorrelated for $G^{(2)}(d)=0$. If a particle moves a distance $d$, the sites become anti-correlated, and the correlator decreases to $G^{(2)}(d)<0$. 

We measure the density-density correlations $G^{2}(d)$ for different disorder strengths in the stationary regime (Fig.~3A). For low disorder, we find the correlations to be independent of distance and below zero. This indicates that the particles tunnel across the entire system and hence are delocalized. On the other hand, at strong disorder, only nearby sites show significant correlations, signaling the absence of particle motion across large distances. We thus conclude that the particles are localized. We extract the correlation length by fitting an exponentially decaying function to the data (Fig.~3B) (SI). For increasing disorder, the correlation length decreases from the entire system size down to around one lattice site (Fig.~3C). 

Our observation of localized particles is consistent with the description of MBL in terms of local integrals of motion \cite{Serbyn2013, Serbyn2013a, Huse2014}. This model was initially formulated for MBL in a spin system, but can be extended to lattice bosons. It describes the global eigenstates as product states of exponentially localized orbitals. The correlation length extracted from our data is a measure of the size of these orbitals. Since the latter form a complete set of locally conserved quantities, this picture connects the breakdown of thermalization in MBL with non-thermalizing, integrable systems. 

\section*{Dynamics and spreading of entanglement}

We now turn to a characterization of the entanglement properties of larger subsystems, starting with a subsystem covering half the system size. As for the case of a single lattice site, the particle number in the subsystem can become entangled with the number in the remaining system through tunneling dynamics, resulting in the number entropy $S_\text{n} = -\sum_{n} p_n \log\left( p_n\right)$. However, subsystems which extend over several lattice sites, with a given particle number, offer the particle configuration as an additional degree of freedom for the entanglement. Configurational entanglement only builds up substantially in interacting systems, since configurational correlations require several particles. The associated configurational entropy $S_\text{c}$, together with the number entropy, forms the von Neumann entropy, $S_\text{vN} = S_\text{n} + S_\text{c}$ (SI). An analogous relation exists for spin systems with conserved total magnetization instead of the particle number.

The dynamics of $S_\text{n}$ and $S_\text{c}$ in the MBL regime can be understood in the picture of localized orbitals. Since the localized orbitals restrict the particle motion, the number entropy can only develop within the localization length and hence $S_\text{n}$ saturates at a lower value than for the thermal case. In the MBL regime, disorder suppresses the tunneling. Therefore, saturation is reached at a later time. However, the dynamics of $S_\text{c}$ are strikingly different. The bare on-site interaction and particle tunneling combine into an effective interaction among localized orbitals, which decays exponentially with the distance between them. As a consequence, entanglement between distant orbitals forms slowly, causing a logarithmic growth of $S_\text{c}$, even after $S_\text{n}$ has saturated \cite{Serbyn2013, Serbyn2013a, Huse2014, Znidaric2008, Bardarson2012}.

\begin{figure*}[t]
	\centering
	\includegraphics[scale= 0.94]{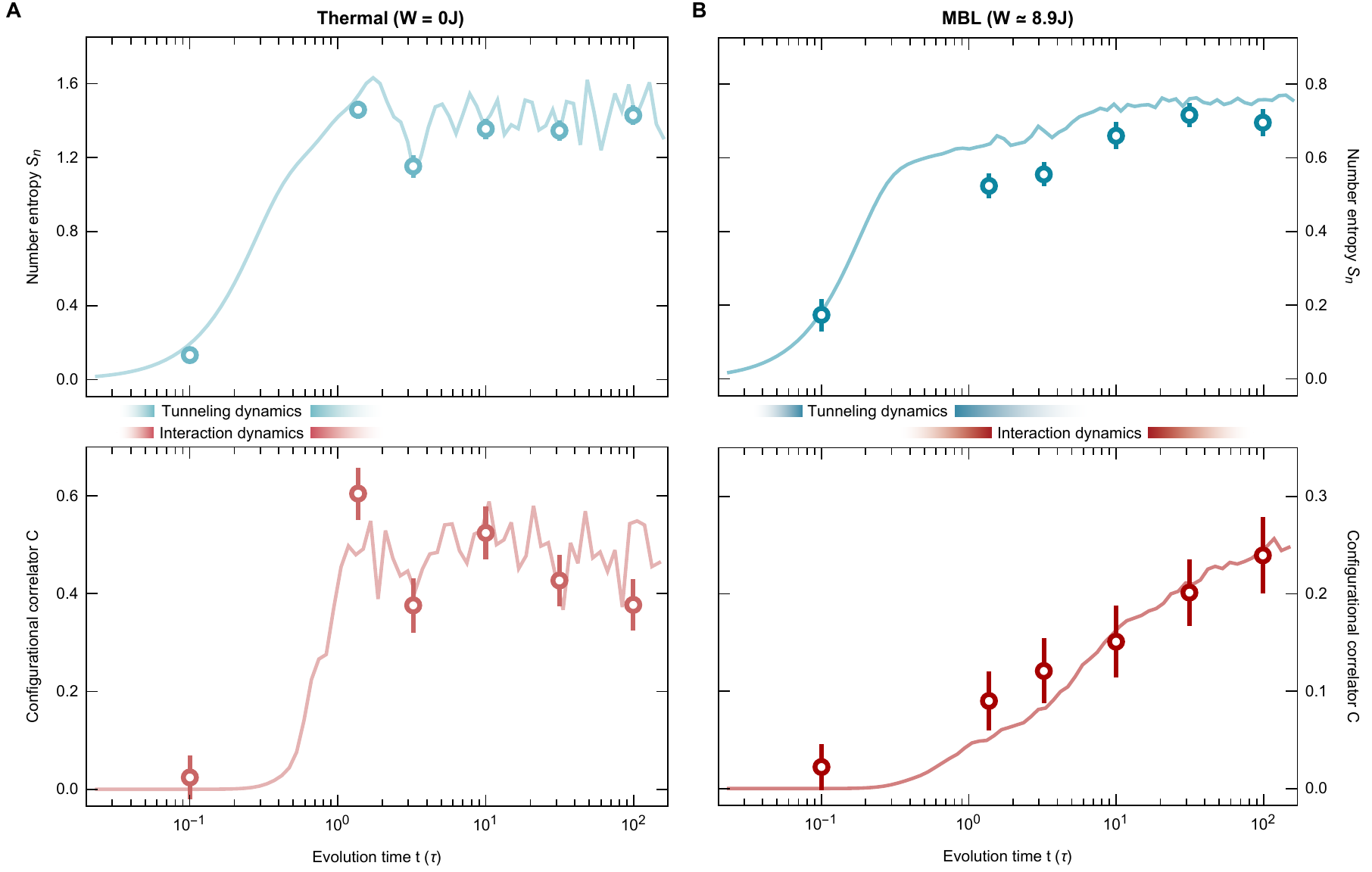}
	\caption{{\bf Dynamics of number and configurational entanglement. (A)} In the thermal regime, both the number entropy $S_\text{n}$ and the configurational correlator $C$ quickly rise and reach a stationary value after thermalization. {\bf (B)} We observe different time scales in the MBL regime. $S_n$ increases for a longer time and reaches a stationary value that is suppressed compared to the thermal one. $C$ shows a persistent slow increase that is consistent with a logarithmic growth, until the longest evolution times covered by our measurements. The solid lines show the prediction of exact diagonalization calculations without any free parameters. The above data was taken on a six-site system and averaged over four disorder realizations. Error bars denote the s.e.m. and are smaller then the markers if hidden.}\label{fig:correlations}
\end{figure*}

In our experiment, we can independently probe both types of entanglement. We obtain the number entropy $S_\text{n}$ through the probabilities $p_n$ by counting the atom number in the subsystem in different experimental realizations. The configurational entropy $S_\text{c}$, in contrast, is challenging to measure in a many-body system since it requires experimental access to the coherences between a large number of quantum states \cite{Islam2015, Elben2017}. Here we choose a complementary approach to probe the configurational entanglement in the system. It exploits the configurational correlations between the subsystems, quantified by the correlator (SI):
\begin{equation}
C = \sum_{n=0}^{N}p_n \sum_{\{A_{n}\},\{B_{n}\}}\left| p(A_{n} \otimes B_{n}) - p(A_{n}) p(B_{n}) \right| ,
\label{eq:C_AB}
\end{equation}
where $\{A_{n}\}$ ($\{B_{n}\}$) is the set of all possible configurations of $n$ particles in subsystem A ($N-n$ in B), and $N$ is total number of particles in the system. All probability distributions are normalized within the subspaces of $n$ particles in A and the remaining $N-n$ particles in B. The configuration $A_{n} \otimes B_{n}$ is separable if $p(A_{n} \otimes B_{n}) = p(A_{n}) p(B_{n})$. The correlator therefore probes the entanglement through the deviation from separability between A and B. In the MBL regime, for sufficiently small amounts of entanglement, we numerically find $C$ to be proportional to $S_\text{c}$, and hence it inherits its scaling properties (SI). Our measurements lie within the numerically verified parameter regime. 

We study the time dynamics of $S_\text{n}$ and $C$ with and without disorder (Fig.~4). Without disorder, both $S_\text{n}$ and $C$ rapidly rise and reach a stationary value within a few tunneling times. In the presence of strong disorder, we find a qualitatively different behavior for both quantities. Again, $S_\text{n}$ reaches a stationary state, although after longer evolution time due to reduced effective tunneling. Additionally the stationary value is significantly reduced, indicating suppressed particle transport through the system. The correlator $C$, in contrast, shows a persistent slow growth up to the longest evolution times reached by our measurements. The growth is consistent with logarithmic behavior over two decades of time evolution. We conclude that we observe interaction-induced dynamics in the MBL regime, which are consistent with the phenomenological model \cite{Serbyn2013, Serbyn2013a, Huse2014}. The agreement of the long-term dynamics of $S_\text{p}$ and $C$ with the numerical calculations in the MBL regime confirms the unitary evolution of the system. 

\begin{figure}[h!]
	\centering
	\includegraphics[scale= 0.94]{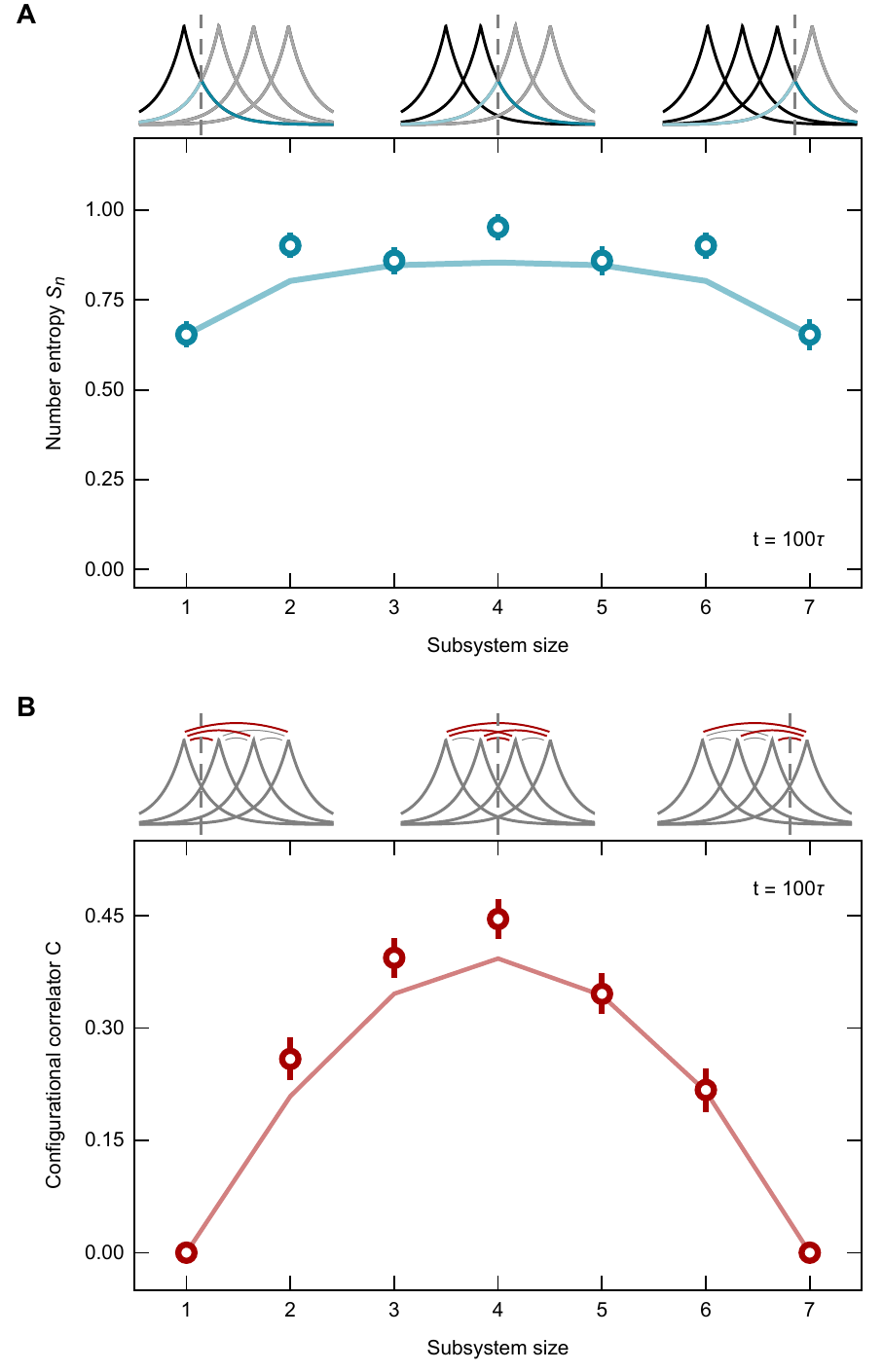}
	\caption{{\bf Spatial distribution of the entanglement.} Number entropy and configurational correlator in the MBL regime ($W=8.9\,J$) after an evolution time of $100\tau$. {\bf (A)} In an MBL system, number fluctuations between two subsystems only stem from local orbitals near the boundary. Consequentially, the number entropy Sn does not depend on the subsystem size, i.e. follows an area law. {\bf (B)} After long evolution times, each local orbital is configurationally entangled with every other. Hence, the configurational correlator C increases almost linearly with the subsystem size, showing a volume-law behavior. The solid lines show the prediction of exact diagonalization calculations without any free parameters. The above data was averaged over four disorder realizations. Error bars denote the s.e.m. and are smaller then the markers if hidden.}
	\label{fig:scalings}
\end{figure}

Considering the entropy in subsystems of different size gives us insights into the spatial distribution of entanglement in the system: in a one-dimensional system, locally generated entanglement results in a subsystem size independent entropy, whereas entanglement from non-local correlations causes the entropy to increase in proportion to the size of the subsystem. In reference to the subsystem's boundary and volume, these scalings are called area law and volume law. We find almost no change in $S_\text{n}$ for different subsystems of an MBL system (Fig.~5A)--- indicating an area law scaling due to localized particles and confirming that particle transport is suppressed. In contrast, the configurational correlations $C$ increase until the subsystem reaches half the system size (Fig.~5B). Such a volume-law scaling is also expected for the entanglement entropy and demonstrates that the observed logarithmic growth indeed stems from non-local correlations across the entire system.

\section*{Conclusion}

Investigating the growth of non-local quantum correlations has been a long-standing experimental challenge for the study of MBL systems. In addition to achieving exceptional isolation from the environment and local access to the system, such a measurement requires access to the entanglement entropy \cite{Islam2015}. Our work provides a novel technique to characterize the entanglement properties of MBL systems, based on measurements of the particle number fluctuations and their configurations. The observation of slow coherent many-body dynamics along with the breakdown of thermalization allows us to unambiguously identify and characterize the MBL state in our system.

In future, experiments at different system sizes will be of interest to shed light on the critical properties of the thermal-to-MBL phase transition, which are the subject of ongoing studies \cite{Vosk2015, Potter2015, Khemani2017, Lueschen2017a}. In our system, it is experimentally feasible to continue scaling the system size at unity filling to a numerically intractable regime. Additionally, we have full control over the disorder potential on every site, which opens the way to studying the role of rare regions and Griffiths dynamics as well as the long-time behavior of an MBL state with a link to a thermal bath \cite{Agarwal2017, Roeck2017, Nandkishore2017}. Ultimately, these studies will further our understanding of quantum thermodynamics and whether such systems are suitable for future applications as quantum memories \cite{Nandkishore2015, Banuls2017}.

\section*{Acknowledgments}   

We acknowledge discussions with I. Cirac, E. Demler, J. Eisert, C. Gross, W. W. Ho, D. A. Huse, H. Pichler, and A. Polkovnikov. We are supported by grants from the National Science Foundation, the Gordon and Betty Moore Foundations EPiQS Initiative, an Air Force Office of Scientific Research MURI program, an Army Research Office MURI program and the NSF Graduate Research Fellowship Program. J. L. acknowledges support from the Swiss National Science Foundation. V.K. is supported by the Harvard Society of Fellows and the William F. Milton Fund.

%


\newpage

\section{Supplementary Information}
\noindent
Supplementary Text\\
Figs. S1 to S10\\
Table S1\\
References \textit{(44-50)}

\makeatletter 
\renewcommand{\thefigure}{S\@arabic\c@figure}
\renewcommand{\theequation}{S\@arabic\c@equation}
\makeatother
\setcounter{figure}{0}  
\setcounter{equation}{0}  

\subsection{Calibration of Bose-Hubbard parameters}

In order to properly frame our experimental results and to compare them to numerical simulations, it is crucial to know the tunneling constant $J$ and on-site interaction strength $U$ in our system. The following two subsections describe how they are measured.

\subsubsection{J calibration}

 In order to calibrate the tunneling constant $J$ in our system, we use two digital micromirror devices (DMDs) to isolate a one-dimensional chain of atoms from the $n=1$ shell of a Mott insulator \cite{Zupancic2016}. With tunneling along the chain being suppressed, we subsequently drop the transverse lattice depth to the value used in our experiments and let the system evolve, causing each atom to perform a one-dimensional quantum walk \cite{Preiss2015}. The probability density distribution describing such a quantum walk is given by 
\begin{equation}\label{eqn:qw}
\left\vert \psi_i(t)\right\vert^2 = \left\vert \mathcal{J}_i \left(\frac{2J}{\pi E}\sin(\pi E t/h)\right)\right\vert^2
\end{equation}
where $i$ denotes the distance to the initial atom position in lattice sites, $\mathcal{J}_i$ is the $i^{th}$-order Bessel function of the first kind, $h$ is Planck's constant, and the nearest-neighbor potential difference $E$ is introduced to allow for an unintentional tilt of the lattice \cite{Hartmann2004}. By averaging over the entire chain of atoms for each image and taking large numbers of images for various evolution times, we are able to measure this density distribution experimentally. The tunneling constant $J$ and lattice tilt $E$ are then determined from a two-parameter fit of equation Eq.~\ref{eqn:qw} to our data. For all experiments discussed in the main publication, the resulting tilt was consistent with a value of $E/h = 0$ Hz. The tunneling was determined to $J/h = 37.5(1)$ Hz.

\subsubsection{U calibration}

In order to calibrate the on-site interaction strength U, we again prepare an $n=1$ Mott insulator. We then use a magnetic field gradient to apply a linear potential along the direction of the (so-far absent) one-dimensional system we are interested in studying. This creates an offset energy $E$ per lattice site. Subsequently, we drop the lattice depth to an intermediate value where tunneling is principally significant but still suppressed by the tilt. We then modulate the depth of our optical lattice and measure the $n=1$ population fraction as a function of the modulation frequency for several regions of interest $i$ (Fig.~\ref{fig:Ucal}a) \cite{Ma2011}. The resulting curves show two minima at energies $E\pm U_i$ (Fig.~\ref{fig:Ucal}b), from which we extract the interaction energy $U_i$. Since our experiments are performed at a lower lattice depth, we rescale the measured interaction energy with $U\propto V^{0.33}_0$. The exponent is determined from numerical calculations that include effects from higher bands. The resulting interaction energy at $V_0=8E_\text{r}$ is determined to be $U/h = 107(1)\,\text{Hz}$.

\subsection{Disorder potential}

In order to study many-body localization, we project a quasiperiodic disorder potential onto our atoms. The following subsections describe the exact shape of this potential, how we calibrate its strength in the plane of the atoms, and how its structure relates to the $G^{(2)}$-measurements discussed in our paper.

\begin{figure}[b]
	\centering
	\includegraphics[width= \columnwidth]{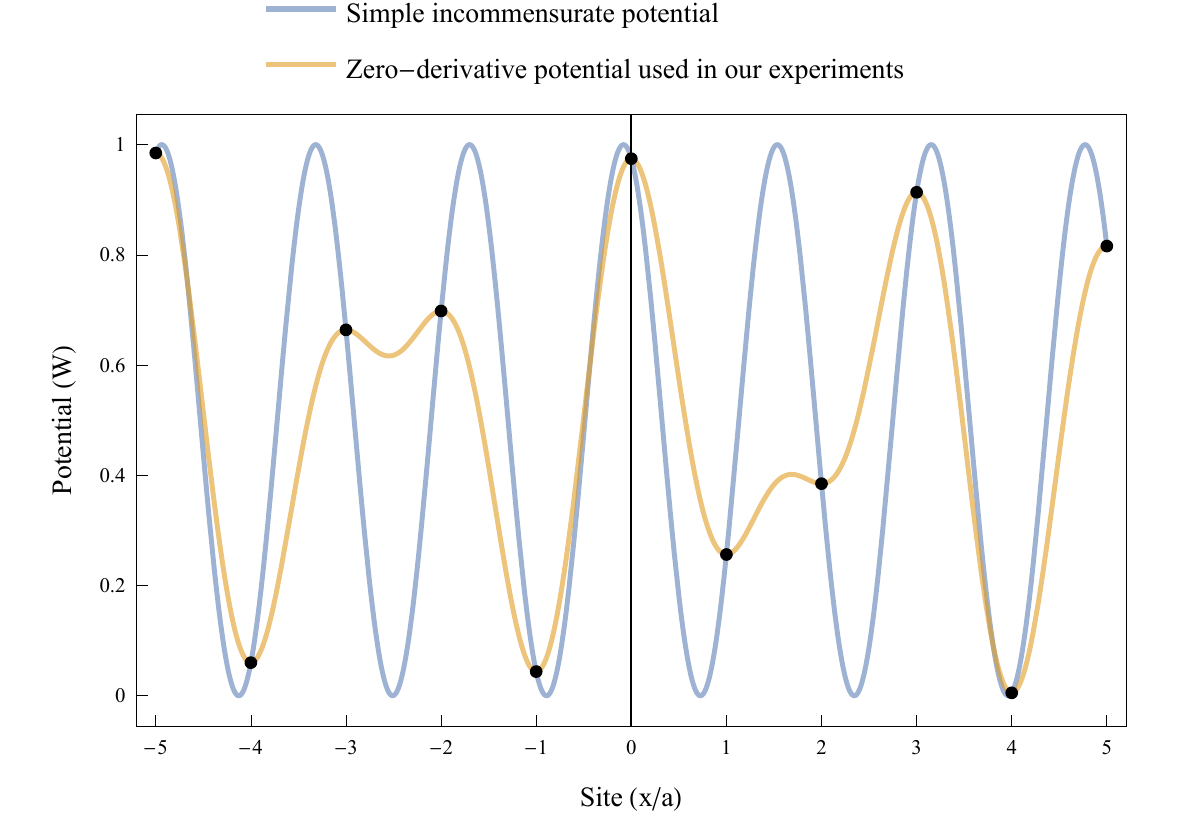}
	\caption{\label{fig:disorder} \textbf{Disorder engineering.} Plot showing a simple incommensurate lattice potential (blue) and the custom potential used in our experiments (yellow). The custom potential was engineered to yield the same on-site values as the simple one, but additionally possesses a vanishing first derivative at the position of the atoms (black dots), thereby making the system less sensitive to shaking-induced heating processes.}
\end{figure}

\subsubsection{Disorder potential generation}

\begin{figure*}[t]
	\includegraphics{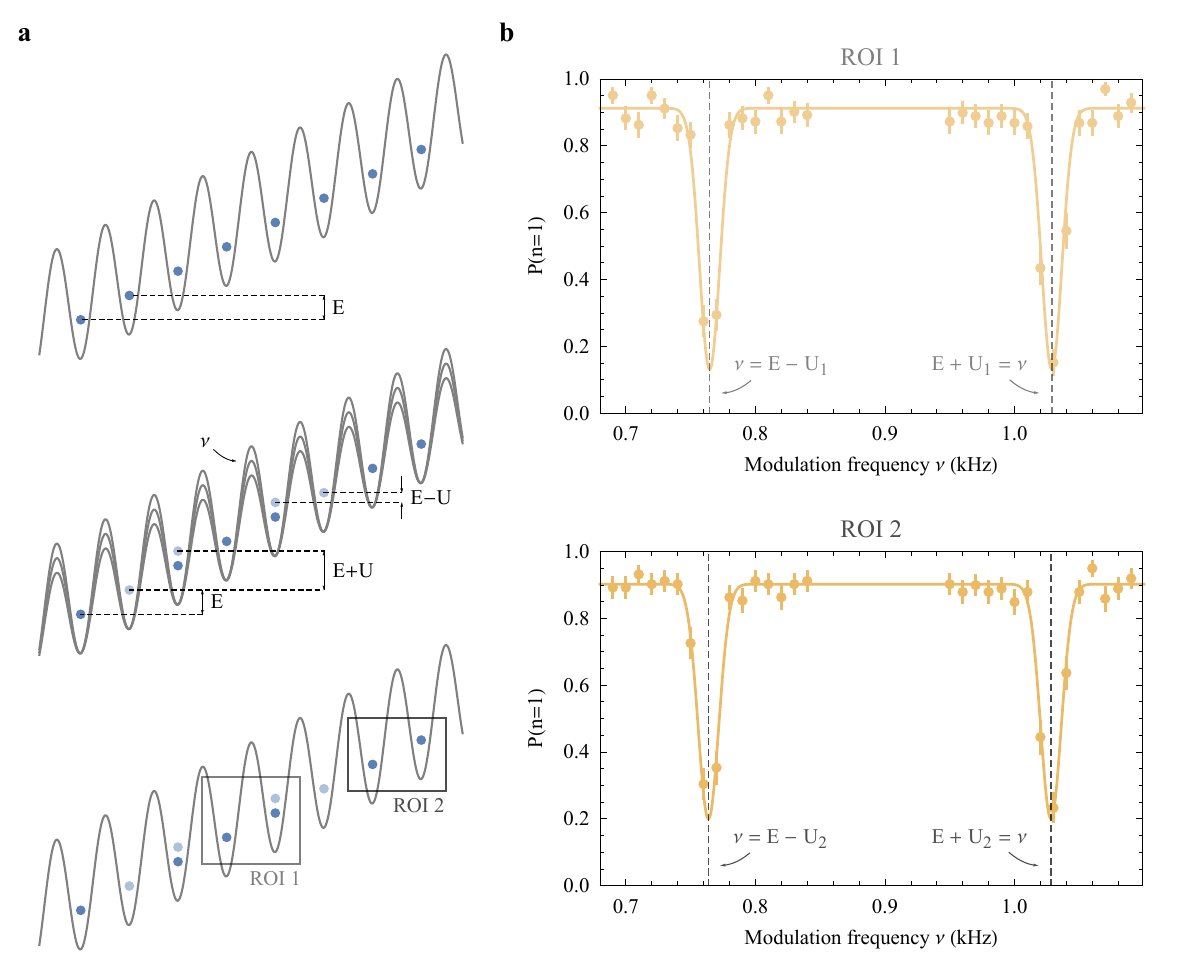}
	\caption{\label{fig:Ucal} \textbf{Measurement of the interaction strength $U$ in the system.} \textbf{(a)} We start with an $n=1$ Mott-insulator in a linear potential and drop the lattice depth to an intermediate value where tunneling is principally allowed but suppressed by the tilt (top). We then modulate the depth of the optical lattice and excite resonances at energies $E\pm U$ depending on the lattice modulation frequency $\nu$ (center). Finally, we image the outcome of the modulation with our quantum gas microscope and obtain number statistics for different regions of interest $i$ (bottom). \textbf{(b)} We post-select on measurement outcomes where a given region contained exactly two atoms and plot the probability of finding them on separate sites. The resulting plots show the expected dips at $E\pm U_i$, from which we obtain the individual values of $U_i$. Averaging over a total of six regions of interest and rescaling the average value to the lattice depth used in the experiment, we determined the interaction strength in our system to equal $U/h = 107(1)$ Hz.} 
\end{figure*}

\begin{figure*}[t]
	\centering
	\includegraphics{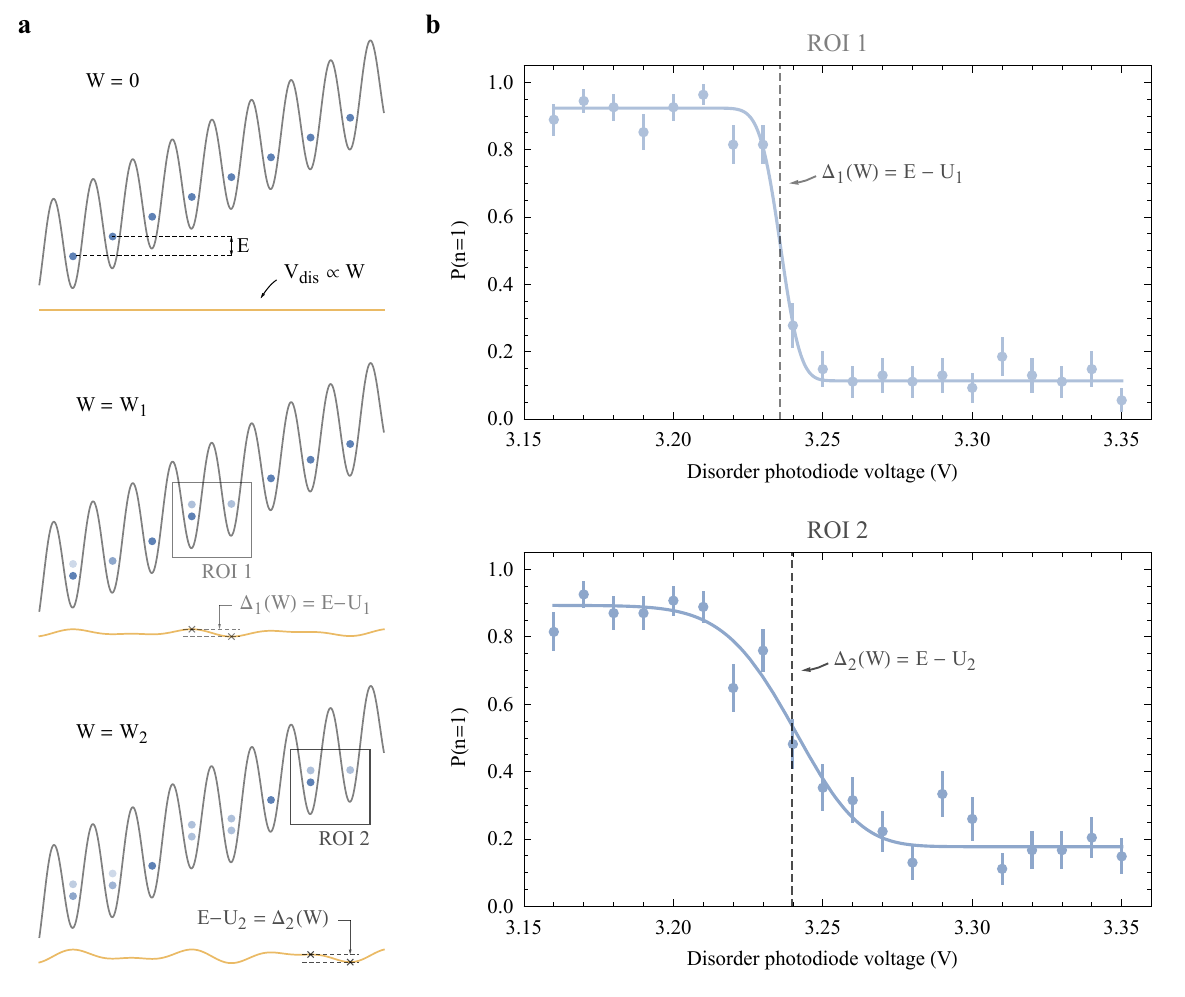}
	\caption{\label{fig:Wcal} \textbf {Calibration of the disorder strength $W$ in our system.}  \textbf{(a)} We prepare the same tilted lattice system as for the $U$ calibration (top). Afterwards, we adiabatically ramp the optical disorder power to different voltages on the corresponding photodiode and analyze the resulting number statistics for various regions of interest. If the disorder power $W$ is made sufficiently large, the potential difference $\Delta_i(W)$ it induces within a given region of interest may partially compensate for the linear tilt and enable resonant tunneling processes (center, bottom). \textbf{(b)} We again post-select on measurement outcomes with exactly two atoms in a given region of interest and plot the probability $P(n=1)$ of finding them on separate sites. The resulting curves show a sharp decay around the photodiode voltages where $\Delta_i(W) = E-U_i$, a condition allowing for resonant tunneling processes that deplete the initial $n=1$ population. Having calibrated the $E-U_i$ values independently and knowing the constant ratios $\Delta_i(W)/W$ from the disorder potential we project, we can link the corresponding phototiode voltages to the appropriate values of $W$ in the plane of our atoms. Technically, data from a single region of interest would be enough to calibrate $W$. The independent calibration of various regions additionally allows us to verify that the potential seen by the atoms indeed obeys the shape we expect, which we found to be the case in our measurements.}
\end{figure*}

We use a digital micromirror device (DMD) in the Fourier plane of the high-resolution imaging system to project a disorder potential onto the atoms. A conceptually simple way of producing such a potential is to project a second lattice with incommensurate periodicity onto the bare lattice holding the atoms, e.g. by adding a potential of the form

\begin{equation}\label{eqn:Vsimple}
V_\text{simple}(x) = 2W \cos^2\left(\pi\frac{x}{\beta a}+\phi\right)
\end{equation}

to the existing lattice. Here, $\beta = \frac{1+\sqrt{5}}{2}$ is the golden ratio, an irrational number that is chosen to ensure maximal incommensurability between the two lattices.

The potential in Eq.~\ref{eqn:Vsimple} generates the desired on-site potential values with a non-vanishing first derivative at the positions of the individual atoms (Fig.~\ref{fig:disorder}). This implies that the potential values sampled by the atoms may vary if the relative position of the disorder potential with respect to the bare lattice is perturbed, thereby making the system susceptible to shaking-induced heating processes. In order to make the system more robust, we set out to find a potential $V(x)$ that would provide the same on-site potential values as the potential in Eq.~\ref{eqn:Vsimple}, while also possessing a vanishing first derivative at the individual sites of the bare lattice, i.e. a potential $V(x)$ such that

\begin{equation}
\begin{split}
\left.V(x)\right\vert_{x = na} &= \left.V_\text{simple}(x)\right\vert_{x = na} \hspace{1mm} ,\\
\left. \frac{\partial V(x)}{\partial x} \right\vert_{x = na} &= 0 \hspace{4mm} \forall \hspace{1mm} n \in \mathbb{Z}
\end{split}
\end{equation}
where $a$ is the lattice constant of the bare lattice, and the position of an arbitrary lattice site is defined to be $x=0$. Since the incommensurate potential is sampled where $x$ is an integer multiple of $a$, the following holds:

\begin{equation}
\begin{split}
\cos^2\left(\pi\frac{n}{\beta} +\phi\right)=\cos^2\left(\pi n \left(\beta-1 \right) +\phi\right)\\
=\cos^2\left(\pi n \left(\beta-m \right) +\phi\right)\hspace{4mm} \forall \hspace{1mm} n,m \in \mathbb{Z}
\end{split}
\end{equation}
where the first equality is due to $\beta-1=\frac{1}{\beta}$. Therefore, one suitable potential is given by the sum of two incommensurate potentials which are displaced by one lattice site:  

\begin{equation}\label{eqn:Vapplied}
\begin{split}
	V_\text{dis}(x) = & 2 W\left [(2-\beta)\cos^2 \left(\pi(\beta-1)\frac{x}{a} + \phi \right) \right. \\
	&\left. + (\beta-1)\cos^2 \left(\pi(\beta-2)\frac{x}{a} + \phi \right) \right]
\end{split}
\end{equation}
The potentials applied during our experiments are of the form Eq.~\ref{eqn:Vapplied}. In the above, different disorder strengths correspond to different values of $W$ and different disorder realizations correspond to different values of the phase $\phi$. Fig.~\ref{fig:disorder} shows the potentials Eq.~\ref{eqn:Vsimple} and Eq.~\ref{eqn:Vapplied} in units of $W$ for $\phi = 0.16$.

\subsubsection{Calibration of the applied disorder potential}

\begin{figure}
	\includegraphics[width= \columnwidth]{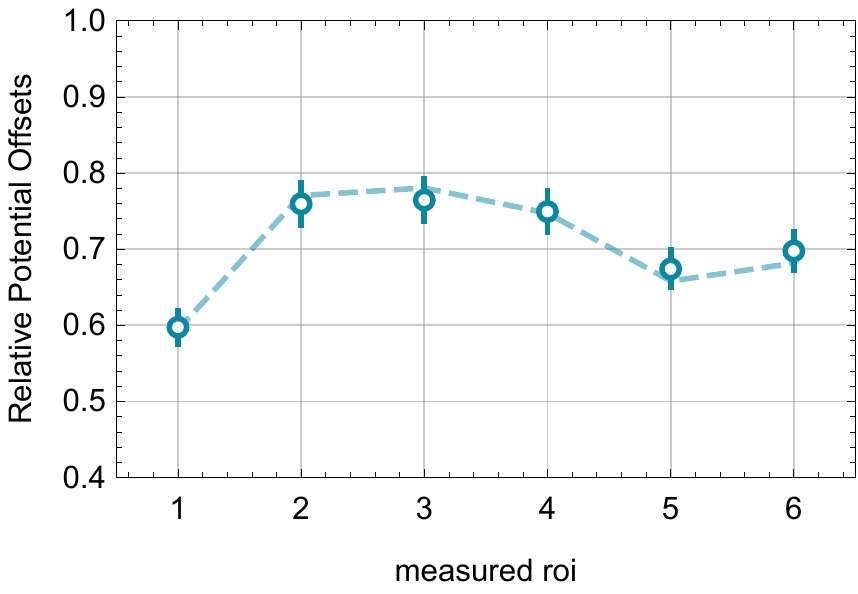}
	\caption{\label{fig:roi_cal} \textbf {Relative potential offsets within one pattern.} In order to benchmark the applied disorder potential, we compare the measured site offsets within one pattern to the numerical Fourier transform of the corresponding hologram (dotted line). The agreement between the data and numerical prediction confirms the high level of control over the applied optical potential. The relative potential offsets (both data and numerical prediction) are normalized to the largest potential difference in all applied patterns.}
\end{figure}

In order to calibrate the effective disorder strength $W$, we prepare the exact same system as for the $U$ calibration, i.e. a tilted $n=1$ Mott insulator. Instead of modulating the depth of the optical lattice, we adiabatically ramp up the optical power on the disorder-DMD to various values. We then study the resulting atom number statistics in the two-site regions of interest that were already used for the $U$ calibration (Fig.~\ref{fig:Wcal}a). By plotting the post-selected unity-filling fraction for a given region $i$, we can identify the optical power at which the disorder successfully compensates for the known $E-U_i$ detuning induced by the tilt (Fig.~\ref{fig:Wcal}b), thereby allowing for resonant tunneling events that deplete the $n=1$ population of the corresponding lattice sites. In theory, measuring one such resonance for a single disorder pattern would suffice to calibrate $W$, as the relative disorder-strength between the corresponding lattice sites with respect to $W$ is known from Eq.~\ref{eqn:Vapplied}. In practice, however, we independently measure the resonance for each region of interest within a given pattern (Fig.~\ref{fig:roi_cal}). This allows us to verify that equation Eq.~\ref{eqn:Vapplied} is indeed a valid description of the disorder potential seen by the atoms, and that our method is not suffering from significant aberrations or other unwanted distortions. 

\subsubsection{$G_2(d)$ fitting and correlations from quasiperiodic potential }

The two-point, density-density correlation function $G_2(d)$, as defined in Eq.~\ref{eqn:g2}, measures the correlated fluctuations of the wavefunction between sites which are spaced a distance $d$ apart.

\begin{equation}\label{eqn:g2}
G_{2}(d)= \langle \hat{a}^\dagger_i  \hat{a}^\dagger_{i+d}  \hat{a}_i \hat{a}_{i+d} \rangle_{i} - \langle \hat{a}^\dagger_i  \hat{a}_i \rangle_{i} \langle \hat{a}^\dagger_{i+d} \hat{a}_{i+d} \rangle_{i}
\end{equation}

\begin{figure*}[t]
	\centering
	\includegraphics[width=6.9in]{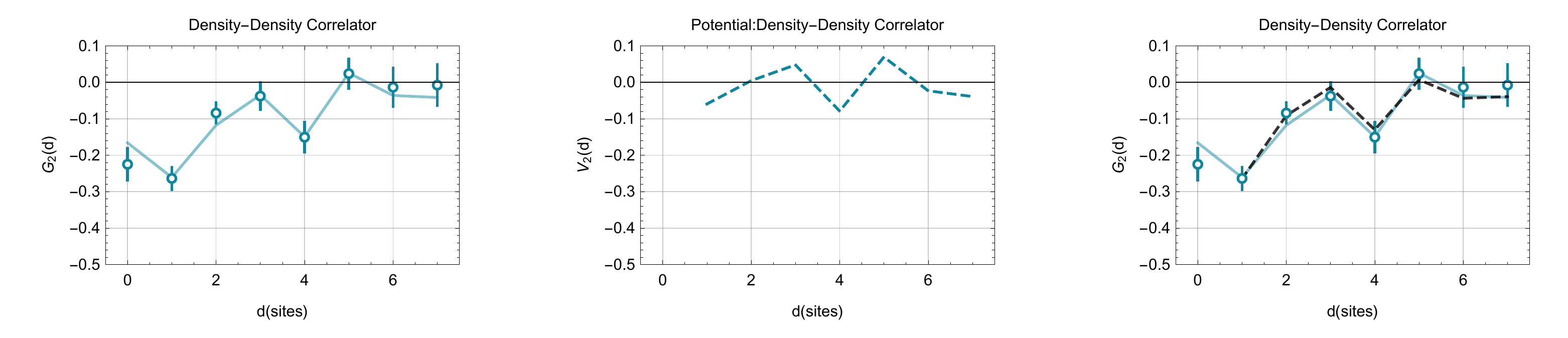}
	\caption{\label{fig:V2row} \textbf{Potential and Wavefunction: Two-Point Correlations} The left plot shows the measured $G_2(d)$ data for an 8 site system after $100\tau$ with disorder strength $W=5.8J$ and $U=2.7J$. The solid line shows the result of exact diagonalization for the same experimental parameters. The middle plot is the calculated two-point correlator for the applied quasiperiodic potential, averaged over many different disorder realizations (Eq.~\ref{eqn:v2}) -- the qualitative features in the left plot mimic the correlation function of the applied potential. The right plot is the same data with the fitted correlation function (black) made from the applied-potential correlations ansatz (Eq.~\ref{eqn:g2fit}).}
\end{figure*}


Due to the conserved global particle number, tunneling alone will generally give rise to an anti-correlation of the $G_2(d)$ function, yielding $G_2(d)<0$. A positive correlation $G_2(d)>0$, on the other hand, corresponds to the wavefunction having density fluctuations which make atoms more likely to be found together at sites $i$ and $i+d$.  For moderate disorder strengths, this two-point correlator obtains a structure which mimics the correlation function of the applied disorder potential (Fig.~\ref{fig:V2row}). The latter is calculated as 
\begin{equation}\label{eqn:v2}
\begin{split}
V_{2}(d)=& \langle V(x,\phi) V(x+d,\phi) \rangle_{d,\phi} \\
&-  \langle V(x,\phi)  \rangle_{d,\phi}  \langle V(x+d,\phi) \rangle_{d,\phi}
\end{split} 
\end{equation}
where $V(x,\phi) = \cos{\left( 2 \pi \beta x + \phi \right) }$ is the applied potential (as sampled by the bare lattice sites).

The similarity between the potential correlation function and the density-density correlation function leads to an ansatz fitting function that captures the non-monotonicity of the density-density correlator:
\begin{equation}\label{eqn:g2fit}
\tilde{G}^{(2)}(d)= \left[A+B \cdot  V_2(d)\right] e^{-d/\xi}
\end{equation}
This function allows for an offset $A$ and rescaling $B$ of the applied-disorder correlations as well as a decay length $\xi$, which is needed to describe localized wavefunctions. This ansatz provides a significantly improved fit for nearly the entire range of measured disordered values. An example is plotted in Fig. \ref{fig:V2row}.

\subsection{Experimental sequence}

\begin{figure*}[t]
	\centering
	\includegraphics{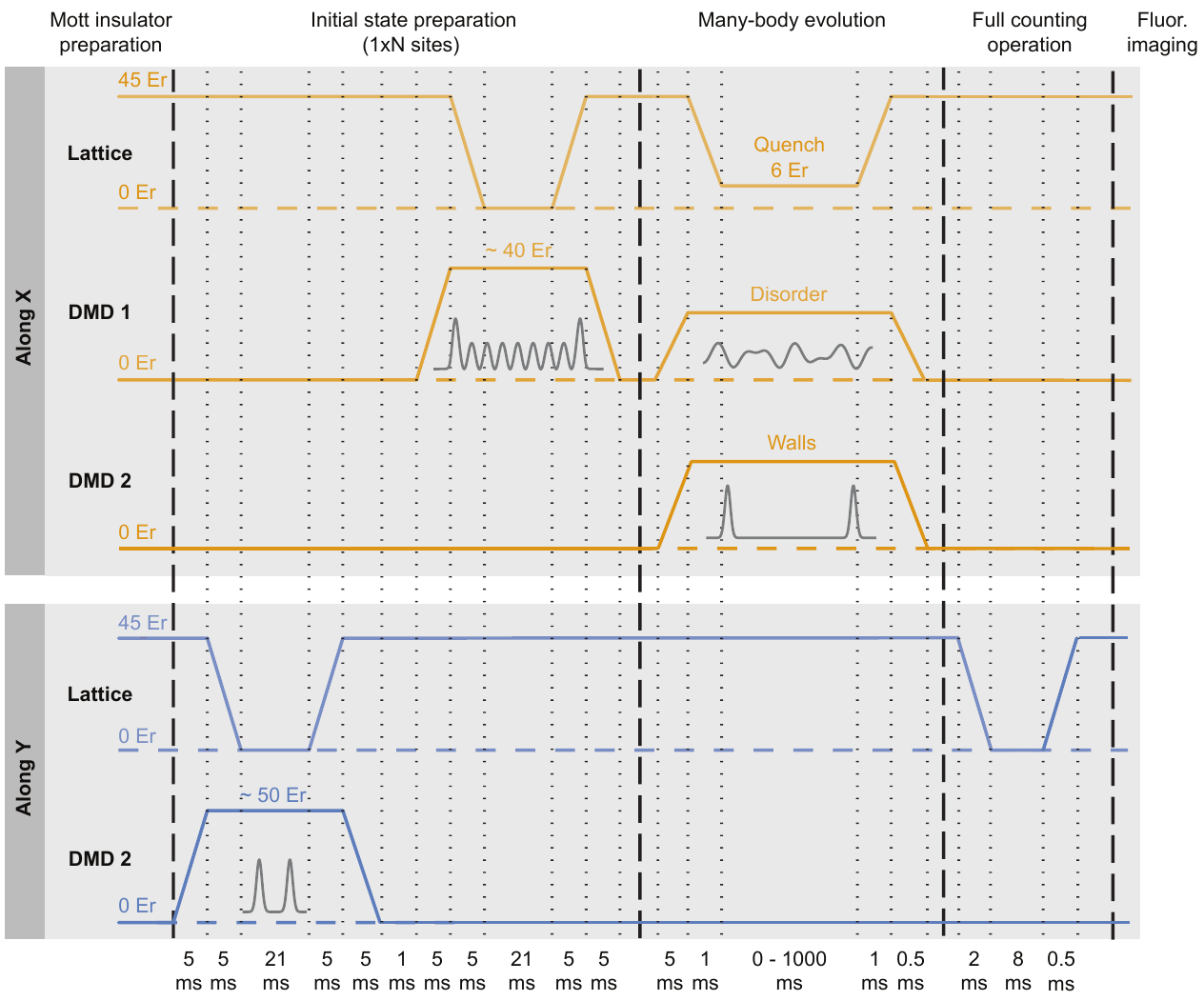}
	\caption{\textbf {Experimental sequence.} Figure showing the sequence of optical potentials used in the experiments. In a two-step process, a $1\times N$ ($N = 6,8$) plaquette of atoms is cut out of an initially prepared Mott insulator, using custom potentials engineered with two different DMDs. Subsequently, one of the DMDs is employed to project a quasi-periodic disorder potential onto the atomic chain, while the other one is used to define the length of the chain after the corresponding lattice has been quenched to a lower power. Following a variable evolution time, the full lattice potential is restored and both DMD potentials are ramped down. We then turn off the transverse lattice to expand the atomic population of the individual lattice sites in tubes. Finally, we perform fluorescence imaging to obtain the full number statistics of the system.}
\end{figure*}

Our experiments start with a two-dimensional, unity-filling Mott insulator of $^{87}$Rb atoms in a deep, blue-detuned optical lattice ($V_x = V_y = 45\,E_\text{r}$) with a lattice constant of $680\,\text{nm}$. Using a procedure outlined in previous work \cite{Kaufman2016}, we employ two digital micromirror devices (DMDs) in the Fourier plane to initialize a one-dimensional system of $N = 8$ ($N=6$ in Fig.~\ref{fig:correlations}) atoms. After post-selection, we achieve a single-site loading fidelity of 99.1(2)\%, which stems from the initial Mott insulator fidelity. Before post-selection, the likelihood of loading more than $N$ atoms is $<0.5(5)\%$.

To initiate the many-body dynamics, we then perform three separate actions. In a first step, we use one of the DMDs to project an optical potential $V_\text{walls}$ onto the atoms, which consists of a flat-top profile in the $y$-direction and two narrow Gaussian peaks separated by $N+2$ lattice sites in the $x$-direction. This "wall-potential" provides a box-like confinement which is registered to the position of the optical lattice, and defines the size of the one-dimensional system once the bare lattice depth has been lowered. Secondly, we simultaneously use the other DMD to project a custom disorder potential onto the atoms. Finally, after both of these potentials have been turned on, we rapidly lower the bare lattice depth along the atomic chain from  $V_x= 45\,E_\text{r}$ to $8\,E_\text{r}$, thereby quenching the system and initiating the dynamics.

After a variable evolution time in this lowered potential, we again freeze the atomic dynamics by quickly ramping the longitudinal lattice back up to $V_x = 45\,E_\text{r}$. We then turn off both the disorder and the confining potential $V_\text{walls}$ and perform a site-resolved atom number measurement. To avoid the parity projection that is usually induced by light-assisted collisions during the imaging process, we employ the following technique: We briefly drop the transverse lattice potential to $V_y\sim0.2\,E_\text{r}$ while keeping the longitudinal lattice at $V_x = 45\,E_\text{r}$. In the absence of a confining transverse lattice, the atoms are free to leave their initial locations and spread out in tubes. After $8\,\text{ms}$ of free evolution in these tubes, the particles have delocalized over approximately 80 lattice sites, and we pin their positions with a deep imaging lattice. In a last step, we illuminate the atoms with a fluorescence beam to acquire site- and number-resolved images of the system, with the total number of atoms in each tube corresponding to the original population of a particular lattice site \cite{Kaufman2016}. 

The above technique, however, does not completely eliminate the possibility of parity projection, since there is still a finite probability to find multiple atoms on the same site even after expansion. To estimate this effect we note that, in the majority of our experiments the probability of having more than three particles on the same site is $1.3(1)\%$. Given the parameters of our expansion, the probability of having a parity projected outcome, starting with 4 atoms on the same site, is estimated to be $4.8(4)\%$. This results in $0.06(3)\%$ of faulty post selected shots, since this effect is smaller then our statistical error we don't account for this it in any of the data.  

\subsection{Thermal ensemble calculation}
The thermal prediction shown in Fig.~\ref{fig:dynamics}E is calculated from an equal-probability statistical mixture of those 11 eigenstates of the Hamiltonian $\hat{\mathcal{H}}$ that are closest to the average energy of the initial state $\ket{\psi_0}$, which is given by $E_0=\bra{\psi_0}\hat{\mathcal{H}}\ket{\psi_0}$. We verify that the results do not depend on the exact number of included eigenstates in the vicinity of the chosen value.

\subsection{Entropy partitioning}

\begin{figure}[t]
	\centering
	\includegraphics[width=\columnwidth]{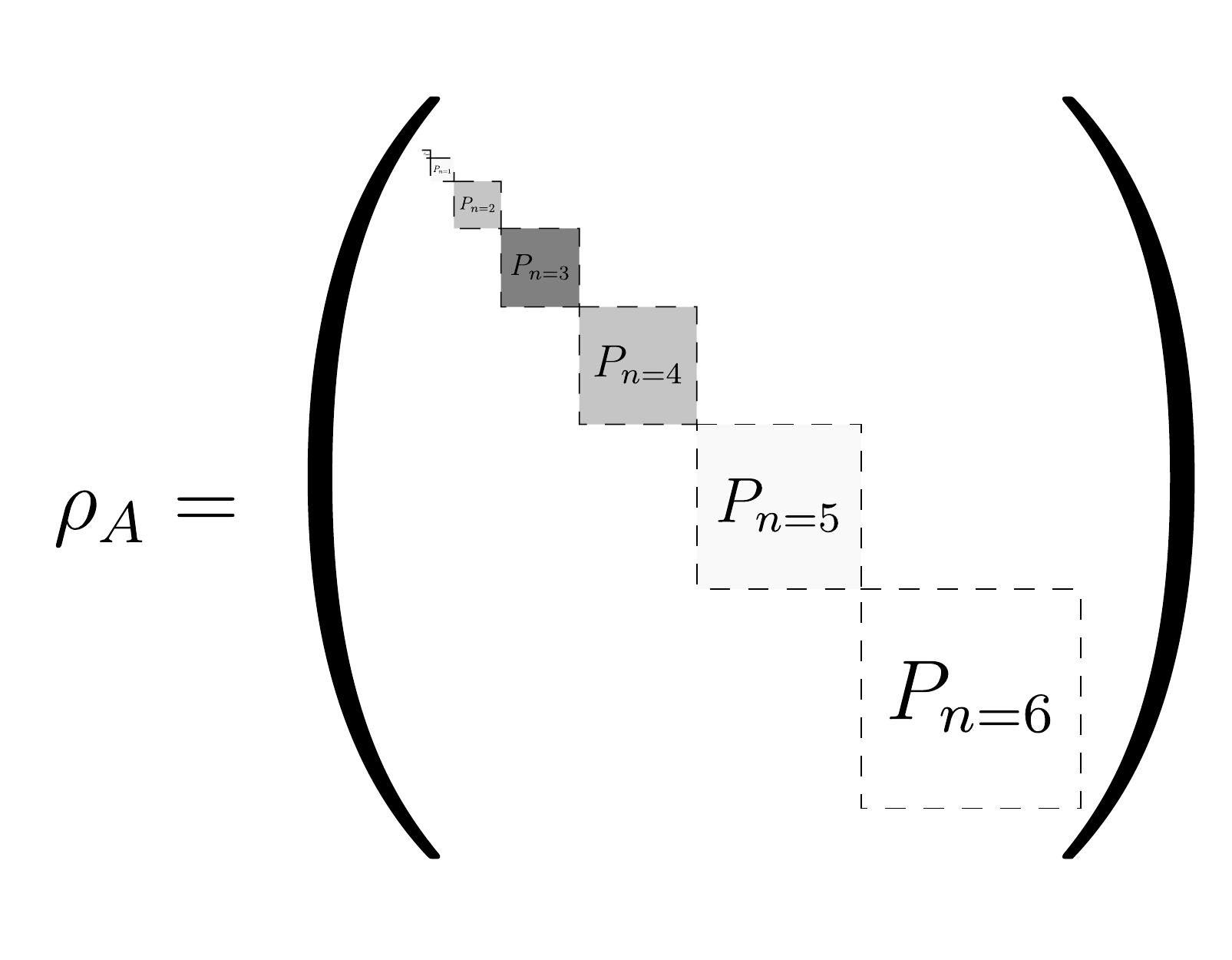}
	\caption{\label{fig:rhoa_pneq} \textbf{Reduced density matrix.} Cartoon of the reduced density matrix with a globally conserved quantity. In the Bose-Hubbard system investigated, the global particle number is conserved: In this example the total particle number is fixed to 6 atoms. The shading in the blocks reflects the relative probability of having that number of atoms in subsystem $A$. The block sizes reflect the relative Hilbert space size corresponding to that number of particles in the subsystem.}
\end{figure}

In a system with a globally conserved quantity, the resultant reduced density matrix for a subsystem is block diagonal. This block diagonal structure is due to the blocks having an exact correlation between two subsystems local quantities that are globally conserved. A specific example of 6 atoms on a Bose-Hubbard chain is shown graphically in Fig.~\ref{fig:rhoa_pneq}. 

The von Neumann entropy for the reduced density matrix $\rho_A$ of subsystem $A$ is defined in the Schmidt basis as
\begin{equation}\label{eqn:Svn}
S_\text{vN}=\sum_i \rho_{ii} \log {\left ( \rho_{ii} \right )},
\end{equation}
Due to the block diagonal structure, $S_\text{vN}$ can be written as the sum of diagonalized blocks
\begin{equation}\label{eqn:Svn2}
S_\text{vN}=\sum_{n=0}^N \sum_i p_n  \rho_{ii}^{(n)} \log {\left (p_n \rho_{ii}^{(n)} \right )},
\end{equation}
where $p_n$ refers to the probability of populating states with $n$ atoms in subsystem $A$ and each block in the reduced density matrix that consists of $n$ atoms is denoted as $\rho^{(n)}$ and normalized according to $\sum_i \rho_{ii}^{(n)} = 1$
The normalized blocks $\rho^{(n)}$ are multiplied by their relative particle number probability in the reduced density matrix by $p_n$. The expression for the von Neumann entropy can then be reduced to a sum of separate entropy contributions $S_\text{n}$ and $S_\text{c}$
\begin{equation}
\begin{aligned}
S_\text{vN}&=\sum_{n=0}^N \sum_i p_n  \rho_{ii}^{(n)} \left[\log{\left ( p_n \right )} + \log {\left ( \rho_{ii}^{(n)} \right )} \right] \\
&=\sum_{n=0}^N p_n  \log{\left ( p_n \right )}  \sum_i \rho_{ii}^{(n)} + \sum_{n=0}^N p_n \sum_{i} \rho_{ii}^{(n)} \log {\left ( \rho_{ii}^{(n)} \right )} \\
&=\sum_{n=0}^N p_n  \log{\left ( p_n \right )}  + \sum_{n=0}^N p_n \sum_{i} \rho_{ii}^{(n)} \log {\left ( \rho_{ii}^{(n)} \right )} \\
&=S_\text{n} + S_\text{c}.
\end{aligned}
\end{equation}

The first term describes the entropy $S_\text{n}$ attributed to the particle number fluctuations within a subsystem and is exactly correlated with the remaining subsystem due to the conserved global particle number. The second term defines the entropy $S_\text{c}$ attributed to the population of different configurational states of $n$ atoms in subsystem $A$. These configurations are internal degrees of freedom for the subsystem $A$, whose Hilbert space dimension depends on the number of atoms $n$ that constrain the possible number of configurations. This type of entanglement has also been refered to as "entanglement of particles'' or "operational entanglement'' \cite{Melko2016, Wiseman2003}.

$S_\text{n}$ is a consequence of the conserved particle number and the hopping present in the system. It will generically have a non-zero contribution for a finite tunnel coupling between subsystems $A$ and $B$. However, $S_\text{c}$ describes the entropy induced by correlations between different configurational states in subsystems $A$ and $B$. These types of correlations are not generically enforced by the mere presence of conserved quantities and dynamics in the system. 

\subsection{Entanglement dynamics in the MBL regime}

The following subsections show exact diagonalization calculations based on the Hamiltonian in Eq.~\ref{eq:H_AA} at strong disorder for both the interacting (MBL) and non-interacting (AL) cases. We numerically show the evolution of the entanglement entropy $S_\text{vN}$ for different Hamiltonian parameters and discuss the connection of the configurational entanglement entropy $S_\text{c}$ with the configurational correlator $C$.

\subsubsection{Entropy partitioning}
\begin{figure}[t]
	\centering
	\includegraphics[width=\columnwidth]{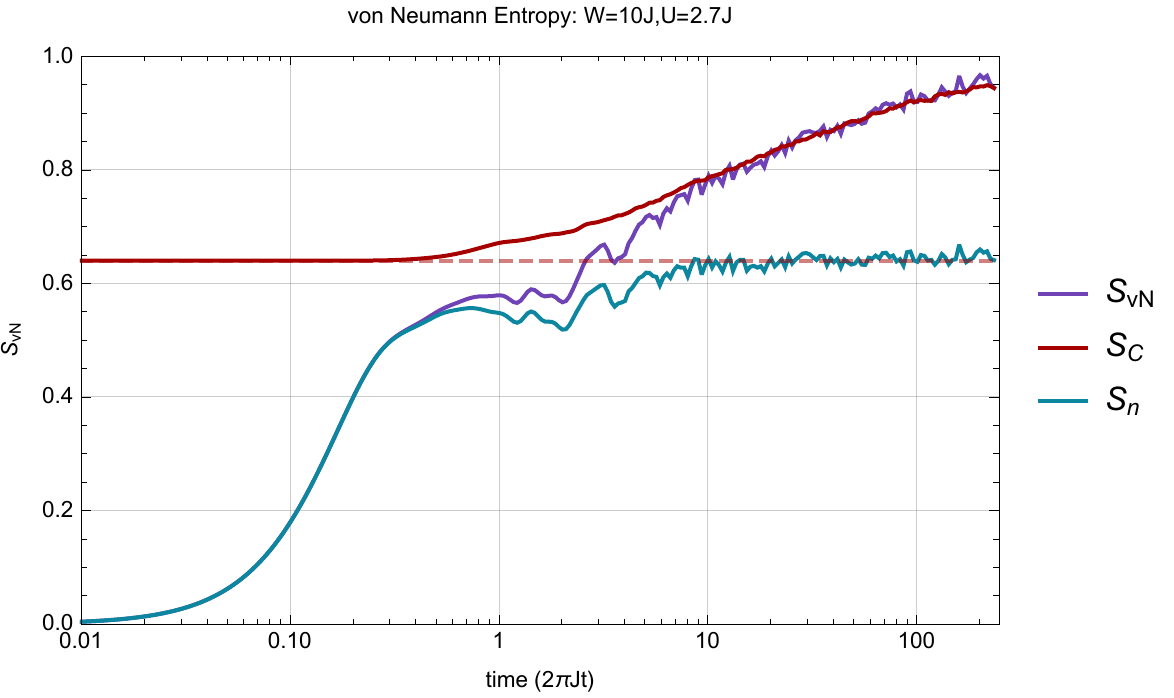}
	\caption{\label{fig:svn_tot} \textbf{Total entropy partitioned} The total von Neumann entanglement entropy $S_\text{vN}$ for the half-system is shown as a function of time in an interacting system at strong disorder. The entropy is split up into $S_\text{n}$ and $S_\text{c}$. For visual guidance, the configurational entropy ($S_\text{c}$) is offset by the long-time average of $S_\text{n}$. This partitioning of the entropy qualitatively shows that logarithmic entropy growth arises primarily from the configurational entropy. The simulation was performed using exact diagonalization on 6 sites at unity filling.}
\end{figure}

A calculation of the dynamics for the partitioned entropy is shown in Fig.~\ref{fig:svn_tot}. Since the initial state is a product state, there is no initial entropy contribution from number fluctuations or configurational correlations. In the many-body-localized regime, the site occupation numbers become a reasonable proxy for locally-conserved quantities, leading to a suppression of the entropy $S_\text{n}$ that is associated with particle fluctuations across the boundary of subsystems $A$ and $B$. Indeed, the numerical calculations show that $S_\text{n}$ reaches a stationary value within a few tunneling times, indicating that the particle transport has reached its equilibrium.

However, the configurational entropy $S_\text{c}$ still grows due to the presence of interactions. This is a consequence of exponentially weak interactions between particles that occupy localized orbitals in each of the subsystems. $S\text{c}$ is responsible for the unbounded logarithmic entropy growth expected in many-body localization \cite{Bardarson2012}. The growth persists for much longer times than the particle number fluctuations and demonstrates a separation of time scales of $S_\text{n}$ and $S_\text{c}$. 

\begin{figure*}[t]
	\centering
	\includegraphics[width=6.9in]{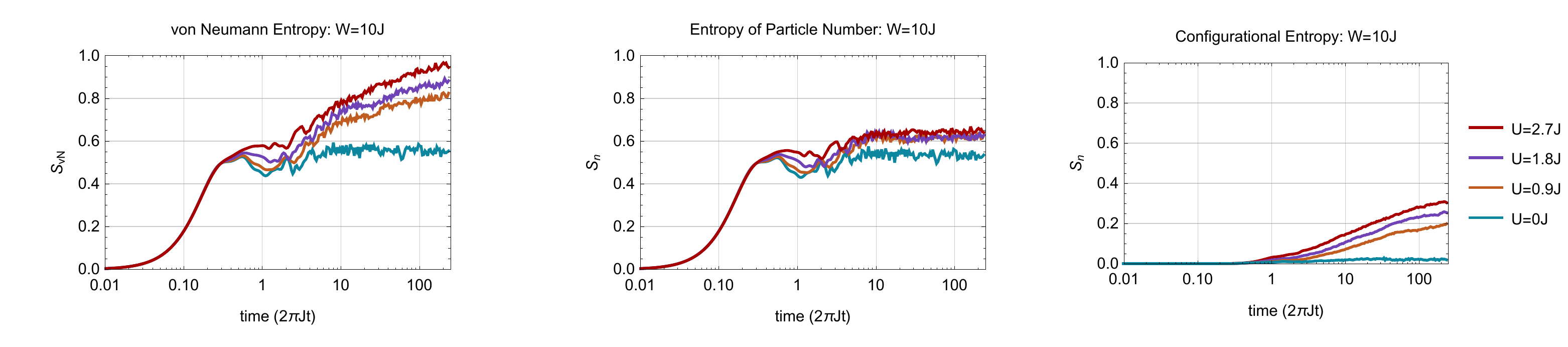}
	\caption{\label{fig:svn_row} \textbf{Entropy comparison: various interaction strengths} The leftmost panel shows the total von Neumann entropy for interaction strengths of $U/J=0,0.9,1.8,2.7$. The second figure shows the contribution of the particle number entropy $S_\text{n}$. The right panel shows the contribution of the configurational entropy $S_\text{c}$ and its dependence on the interaction strength. All simulations were performed by exact diagonalization on 6 sites at unity filling.}
\end{figure*}

\subsubsection{Effect of interactions}
In order to separately investigate the contribution of $S_\text{n}$ and $S_\text{c}$, we perform numerical calculations at different interaction strengths $U$. The total (von Neumann) entanglement entropy $S_\text{vN}$ shows a logarithmically-slow growth, which depends on the interaction strength. Whereas the number entropy $S_\text{n}$ is almost independent of the interaction strength, the configurational entropy $S_\text{c}$ shows a qualitatively different behaviour in the presence or absence of interactions in the system. Without interactions, almost no configurational entropy is generated and $S_\text{c}$ remains nearly independent of the evolution time. However,  when interactions are present, $S_\text{c}$ shows a logarithmic growth with the same interaction dependence as $S_\text{vN}$. These calculations show that the separation of the entanglement into number and configurational degrees of freedom allows us to isolate the logarithmic growth of $S_\text{vN}$. 


\subsubsection{Configurational correlations vs. configurational entropy}

Since the configurational entropy is inaccessible in this experiment, we use the correlator $C$ as a measure of the configurational entanglement in the system (see Eq.~\ref{eq:C_AB}). It is related to the configurational mutual information between the two subsystems \cite{Wolf2008}. $C$ measures the distance between the joint distribution of particle configurations in the entire system from the uncorrelated distributions of configurations in subsystem $A$ and $B$. These correlations are measured in the Fock basis and act as a proxy for the corresponding configurational-entropy growth.

\begin{figure*}[t]
	\centering
	\includegraphics[width=6.9in]{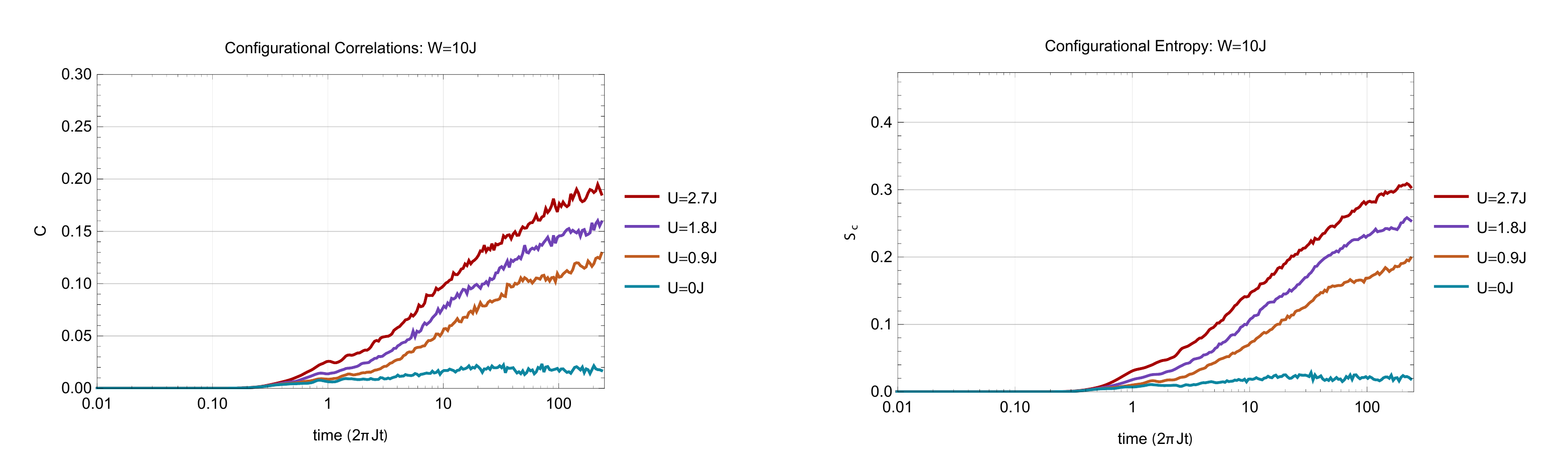}
	\caption{\label{fig:CABvsSc} \textbf{Correlations and Configurational Entropy.} The configurational correlations $C$ in the Fock basis are plotted in the left panel as a function of interaction strength and evolution time. The corresponding configurational entropy $S_\text{c}$ is plotted in the right panel as a function of interaction strength and evolution time. All simulations were performed by exact diagonalization on 6 sites at unity filling.}
\end{figure*}

Despite being qualitatively similar metrics which go to zero in the unentangled limit, $S_\text{c}$ and $C$ show very distinct behavior for large degrees of entanglement. Consider the maximally mixed reduced density matrix in the Schmidt basis. The configurational entanglement entropy $S_\text{c}$ is unbounded and $S_\text{c}\rightarrow \log{(\text{N})}$ for the maximally mixed case with $N$ representing the Hilbert space dimension of the reduced density matrix. 

Since
\begin{equation}
\begin{aligned}
C &= \sum_{n=0}^N p_n \sum_{\{A_n\},\{B_n\}} \left | p(A_{n} \otimes B_{n}) + p(A_{n}) p(B_{n}) \right |\\
 &\leq\sum_{n=0}^N p_n \sum_{\{A_n\},\{B_n\}} \left | p(A_{n} \otimes B_{n})\right | + \left | p(A_{n}) p(B_{n}) \right | = 2
\end{aligned}
\end{equation}
where the last equality is enforced by the normalization of the probability distributions. This bound of $C\leq2$ can be shown to be a tight upper bound in large Hilbert space dimensions for a maximally mixed reduced density matrix, where the two probability distributions are perfectly correlated. Let us consider the maximally mixed reduced density matrix in the Schmidt basis, with $p^{(n)}_a,p^{(n)}_b = 1/N$ and $p^{(n)}_{a,b} \in \{1/N, 0\}$. For simplicity, let us consider the case where $p_n=1$ for one $n_0$ and $p_n=0$ for all other $n$. We therefore drop all superscripts of $n$. 
\begin{equation}
\begin{aligned}
C &= \sum_{\{A\},\{B\}} \left| p(A\otimes B) - p(A) p(B) \right|  \\
&\leq \sum_{a,b = 1}^N \left |\frac{1}{N} - \frac{1}{N^2} \right | + \sum_{a,b = N+1}^{N^2} \left | 0 - \frac{1}{N^2} \right | \\ 
&= N \left (\frac{1}{N} \right ) \left | 1 - \frac{1}{N} \right | +\left ( N^2 - N \right ) \frac{1}{N^2} \\
&= \frac{N-1}{N} + 1 - \frac{1}{N} = 2 \left (1-\frac{1}{N} \right)
\end{aligned}
\end{equation}
This shows that for the same reduced density matrices, $S_\text{c} \rightarrow \log{\text{N}}$ as $C \rightarrow 2 (1-\frac{1}{N})$. While the asymptotic behaviors are different for $S_\text{c}$ and $C$, the performed measurements are far away from this differentiating asymptotic regime, and the values of both quantities are approximately linearly related for the measured parameter range (as numerically shown in Fig. \ref{fig:CABvsSc}).

\begin{table*}[t]
	\centering
	\begin{tabular}{|c|c|c|}
		\hline Figure & number of unique disorders & post selected shots per point\\
		\hline 2D: W=1.0J & 197 & 64, 82, 68, 63, 58, 55, 133, 142, 192, 180, 160, 108\\ 
		\hline 2D: W=8.9J & 197 & 134, 135, 143, 140, 156, 152, 144, 126, 123, 128, 126, 147 \\ 
		\hline 2E & 197 & 106, 113, 118, 129, 191, 186, 101, 186, 185, 104, 110, 124, 102 \\ 
		\hline 3A,B: W=1.0J & 197 & 160\\ 
		\hline 3A,B: W=8.9J & 197 & 110 \\ 
		\hline 3C & 197 & 129, 191, 101, 186, 185, 104, 110, 124, 102 \\ 
		\hline 4 & 4 & avg. per disorder: 210.5, 305.75, 264.25, 219, 263.5, 196.25\\
		\hline 5 & 4 & avg. per disorder: 740.25\\
		\hline
	\end{tabular} 
\end{table*}

\subsection{Data analysis}
For the data in Figs.~\ref{fig:dynamics} and \ref{fig:localization}, we use 197 unique disorder patterns and perform a running average by randomly sampling a given number of realizations and treating them as independent measurements of the same system. Additionally, the data in Fig.~\ref{fig:dynamics} is averaged over only the middle 6 sites of the chain to exclude edge effects. For the data in Figs.~\ref{fig:correlations} and \ref{fig:scalings}, we use 4 different disorder patterns. We first average each observable over different outcomes for the same disorder and subsequently perform the average over different disorder realizations. 


\begin{thebibliography}{50}%
	\makeatletter
	\providecommand \@ifxundefined [1]{%
		\@ifx{#1\undefined}
	}%
	\providecommand \@ifnum [1]{%
		\ifnum #1\expandafter \@firstoftwo
		\else \expandafter \@secondoftwo
		\fi
	}%
	\providecommand \@ifx [1]{%
		\ifx #1\expandafter \@firstoftwo
		\else \expandafter \@secondoftwo
		\fi
	}%
	\providecommand \natexlab [1]{#1}%
	\providecommand \enquote  [1]{``#1''}%
	\providecommand \bibnamefont  [1]{#1}%
	\providecommand \bibfnamefont [1]{#1}%
	\providecommand \citenamefont [1]{#1}%
	\providecommand \href@noop [0]{\@secondoftwo}%
	\providecommand \href [0]{\begingroup \@sanitize@url \@href}%
	\providecommand \@href[1]{\@@startlink{#1}\@@href}%
	\providecommand \@@href[1]{\endgroup#1\@@endlink}%
	\providecommand \@sanitize@url [0]{\catcode `\\12\catcode `\$12\catcode
		`\&12\catcode `\#12\catcode `\^12\catcode `\_12\catcode `\%12\relax}%
	\providecommand \@@startlink[1]{}%
	\providecommand \@@endlink[0]{}%
	\providecommand \url  [0]{\begingroup\@sanitize@url \@url }%
	\providecommand \@url [1]{\endgroup\@href {#1}{\urlprefix }}%
	\providecommand \urlprefix  [0]{URL }%
	\providecommand \Eprint [0]{\href }%
	\providecommand \doibase [0]{http://dx.doi.org/}%
	\providecommand \selectlanguage [0]{\@gobble}%
	\providecommand \bibinfo  [0]{\@secondoftwo}%
	\providecommand \bibfield  [0]{\@secondoftwo}%
	\providecommand \translation [1]{[#1]}%
	\providecommand \BibitemOpen [0]{}%
	\providecommand \bibitemStop [0]{}%
	\providecommand \bibitemNoStop [0]{.\EOS\space}%
	\providecommand \EOS [0]{\spacefactor3000\relax}%
	\providecommand \BibitemShut  [1]{\csname bibitem#1\endcsname}%
	\let\auto@bib@innerbib\@empty
	\bibitem [{\citenamefont {Deutsch}(1991)}]{Deutsch1991}%
	\BibitemOpen
	\bibfield  {author} {\bibinfo {author} {\bibfnamefont {J.~M.}\ \bibnamefont
			{Deutsch}},\ }\href {\doibase 10.1103/PhysRevA.43.2046} {\bibfield  {journal}
		{\bibinfo  {journal} {Physical Review A}\ }\textbf {\bibinfo {volume} {43}},\
		\bibinfo {pages} {2046} (\bibinfo {year} {1991})}\BibitemShut {NoStop}%
	\bibitem [{\citenamefont {Srednicki}(1994)}]{Srednicki1994}%
	\BibitemOpen
	\bibfield  {author} {\bibinfo {author} {\bibfnamefont {M.}~\bibnamefont
			{Srednicki}},\ }\href {\doibase 10.1103/PhysRevE.50.888} {\bibfield
		{journal} {\bibinfo  {journal} {Physical Review E}\ }\textbf {\bibinfo
			{volume} {50}},\ \bibinfo {pages} {888} (\bibinfo {year} {1994})}\BibitemShut
	{NoStop}%
	\bibitem [{\citenamefont {Rigol}\ \emph {et~al.}(2008)\citenamefont {Rigol},
		\citenamefont {Dunjko},\ and\ \citenamefont {Olshanii}}]{Rigol2008}%
	\BibitemOpen
	\bibfield  {author} {\bibinfo {author} {\bibfnamefont {M.}~\bibnamefont
			{Rigol}}, \bibinfo {author} {\bibfnamefont {V.}~\bibnamefont {Dunjko}}, \
		and\ \bibinfo {author} {\bibfnamefont {M.}~\bibnamefont {Olshanii}},\ }\href
	{\doibase 10.1038/nature06838} {\bibfield  {journal} {\bibinfo  {journal}
			{Nature}\ }\textbf {\bibinfo {volume} {452}},\ \bibinfo {pages} {854}
		(\bibinfo {year} {2008})}\BibitemShut {NoStop}%
	\bibitem [{\citenamefont {Neill}\ \emph {et~al.}(2016)\citenamefont {Neill},
		\citenamefont {Roushan}, \citenamefont {Fang}, \citenamefont {Chen},
		\citenamefont {Kolodrubetz}, \citenamefont {Chen}, \citenamefont {Megrant},
		\citenamefont {Barends}, \citenamefont {Campbell}, \citenamefont {Chiaro},
		\citenamefont {Dunsworth}, \citenamefont {Jeffrey}, \citenamefont {Kelly},
		\citenamefont {Mutus}, \citenamefont {O'Malley}, \citenamefont {Quintana},
		\citenamefont {Sank}, \citenamefont {Vainsencher}, \citenamefont {Wenner},
		\citenamefont {White}, \citenamefont {Polkovnikov},\ and\ \citenamefont
		{Martinis}}]{Neill2016}%
	\BibitemOpen
	\bibfield  {author} {\bibinfo {author} {\bibfnamefont {C.}~\bibnamefont
			{Neill}}, \bibinfo {author} {\bibfnamefont {P.}~\bibnamefont {Roushan}},
		\bibinfo {author} {\bibfnamefont {M.}~\bibnamefont {Fang}}, \bibinfo {author}
		{\bibfnamefont {Y.}~\bibnamefont {Chen}}, \bibinfo {author} {\bibfnamefont
			{M.}~\bibnamefont {Kolodrubetz}}, \bibinfo {author} {\bibfnamefont
			{Z.}~\bibnamefont {Chen}}, \bibinfo {author} {\bibfnamefont {A.}~\bibnamefont
			{Megrant}}, \bibinfo {author} {\bibfnamefont {R.}~\bibnamefont {Barends}},
		\bibinfo {author} {\bibfnamefont {B.}~\bibnamefont {Campbell}}, \bibinfo
		{author} {\bibfnamefont {B.}~\bibnamefont {Chiaro}}, \bibinfo {author}
		{\bibfnamefont {A.}~\bibnamefont {Dunsworth}}, \bibinfo {author}
		{\bibfnamefont {E.}~\bibnamefont {Jeffrey}}, \bibinfo {author} {\bibfnamefont
			{J.}~\bibnamefont {Kelly}}, \bibinfo {author} {\bibfnamefont
			{J.}~\bibnamefont {Mutus}}, \bibinfo {author} {\bibfnamefont {P.~J.~J.}\
			\bibnamefont {O'Malley}}, \bibinfo {author} {\bibfnamefont {C.}~\bibnamefont
			{Quintana}}, \bibinfo {author} {\bibfnamefont {D.}~\bibnamefont {Sank}},
		\bibinfo {author} {\bibfnamefont {A.}~\bibnamefont {Vainsencher}}, \bibinfo
		{author} {\bibfnamefont {J.}~\bibnamefont {Wenner}}, \bibinfo {author}
		{\bibfnamefont {T.~C.}\ \bibnamefont {White}}, \bibinfo {author}
		{\bibfnamefont {A.}~\bibnamefont {Polkovnikov}}, \ and\ \bibinfo {author}
		{\bibfnamefont {J.~M.}\ \bibnamefont {Martinis}},\ }\href {\doibase
		10.1038/nphys3830} {\bibfield  {journal} {\bibinfo  {journal} {Nature
				Physics}\ }\textbf {\bibinfo {volume} {12}},\ \bibinfo {pages} {1037}
		(\bibinfo {year} {2016})}\BibitemShut {NoStop}%
	\bibitem [{\citenamefont {Kaufman}\ \emph {et~al.}(2016)\citenamefont
		{Kaufman}, \citenamefont {Tai}, \citenamefont {Lukin}, \citenamefont
		{Rispoli}, \citenamefont {Schittko}, \citenamefont {Preiss},\ and\
		\citenamefont {Greiner}}]{Kaufman2016}%
	\BibitemOpen
	\bibfield  {author} {\bibinfo {author} {\bibfnamefont {A.~M.}\ \bibnamefont
			{Kaufman}}, \bibinfo {author} {\bibfnamefont {M.~E.}\ \bibnamefont {Tai}},
		\bibinfo {author} {\bibfnamefont {A.}~\bibnamefont {Lukin}}, \bibinfo
		{author} {\bibfnamefont {M.}~\bibnamefont {Rispoli}}, \bibinfo {author}
		{\bibfnamefont {R.}~\bibnamefont {Schittko}}, \bibinfo {author}
		{\bibfnamefont {P.~M.}\ \bibnamefont {Preiss}}, \ and\ \bibinfo {author}
		{\bibfnamefont {M.}~\bibnamefont {Greiner}},\ }\href {\doibase
		10.1126/science.aaf6725} {\bibfield  {journal} {\bibinfo  {journal}
			{Science}\ }\textbf {\bibinfo {volume} {353}},\ \bibinfo {pages} {794}
		(\bibinfo {year} {2016})}\BibitemShut {NoStop}%
	\bibitem [{\citenamefont {Nandkishore}\ and\ \citenamefont
		{Huse}(2015)}]{Nandkishore2015}%
	\BibitemOpen
	\bibfield  {author} {\bibinfo {author} {\bibfnamefont {R.}~\bibnamefont
			{Nandkishore}}\ and\ \bibinfo {author} {\bibfnamefont {D.~A.}\ \bibnamefont
			{Huse}},\ }\href {\doibase 10.1146/annurev-conmatphys-031214-014726}
	{\bibfield  {journal} {\bibinfo  {journal} {Annual Review of Condensed Matter
				Physics}\ }\textbf {\bibinfo {volume} {6}},\ \bibinfo {pages} {15} (\bibinfo
		{year} {2015})}\BibitemShut {NoStop}%
	\bibitem [{\citenamefont {Anderson}(1958)}]{Anderson1958}%
	\BibitemOpen
	\bibfield  {author} {\bibinfo {author} {\bibfnamefont {P.~W.}\ \bibnamefont
			{Anderson}},\ }\href {\doibase 10.1103/PhysRev.109.1492} {\bibfield
		{journal} {\bibinfo  {journal} {Physical Review}\ }\textbf {\bibinfo {volume}
			{109}},\ \bibinfo {pages} {1492} (\bibinfo {year} {1958})}\BibitemShut
	{NoStop}%
	\bibitem [{\citenamefont {Wiersma}\ \emph {et~al.}(1997)\citenamefont
		{Wiersma}, \citenamefont {Bartolini}, \citenamefont {Lagendijk},\ and\
		\citenamefont {Righini}}]{Wiersma1997}%
	\BibitemOpen
	\bibfield  {author} {\bibinfo {author} {\bibfnamefont {D.~S.}\ \bibnamefont
			{Wiersma}}, \bibinfo {author} {\bibfnamefont {P.}~\bibnamefont {Bartolini}},
		\bibinfo {author} {\bibfnamefont {A.}~\bibnamefont {Lagendijk}}, \ and\
		\bibinfo {author} {\bibfnamefont {R.}~\bibnamefont {Righini}},\ }\href
	{\doibase 10.1038/37757} {\bibfield  {journal} {\bibinfo  {journal} {Nature}\
		}\textbf {\bibinfo {volume} {390}},\ \bibinfo {pages} {671} (\bibinfo {year}
		{1997})}\BibitemShut {NoStop}%
	\bibitem [{\citenamefont {Schwartz}\ \emph {et~al.}(2007)\citenamefont
		{Schwartz}, \citenamefont {Bartal}, \citenamefont {Fishman},\ and\
		\citenamefont {Segev}}]{Schwartz2007}%
	\BibitemOpen
	\bibfield  {author} {\bibinfo {author} {\bibfnamefont {T.}~\bibnamefont
			{Schwartz}}, \bibinfo {author} {\bibfnamefont {G.}~\bibnamefont {Bartal}},
		\bibinfo {author} {\bibfnamefont {S.}~\bibnamefont {Fishman}}, \ and\
		\bibinfo {author} {\bibfnamefont {M.}~\bibnamefont {Segev}},\ }\href
	{\doibase 10.1038/nature05623} {\bibfield  {journal} {\bibinfo  {journal}
			{Nature}\ }\textbf {\bibinfo {volume} {446}},\ \bibinfo {pages} {52}
		(\bibinfo {year} {2007})}\BibitemShut {NoStop}%
	\bibitem [{\citenamefont {Billy}\ \emph {et~al.}(2008)\citenamefont {Billy},
		\citenamefont {Josse}, \citenamefont {Zuo}, \citenamefont {Bernard},
		\citenamefont {Hambrecht}, \citenamefont {Lugan}, \citenamefont
		{Cl{\'{e}}ment}, \citenamefont {Sanchez-Palencia}, \citenamefont {Bouyer},\
		and\ \citenamefont {Aspect}}]{Billy2008}%
	\BibitemOpen
	\bibfield  {author} {\bibinfo {author} {\bibfnamefont {J.}~\bibnamefont
			{Billy}}, \bibinfo {author} {\bibfnamefont {V.}~\bibnamefont {Josse}},
		\bibinfo {author} {\bibfnamefont {Z.}~\bibnamefont {Zuo}}, \bibinfo {author}
		{\bibfnamefont {A.}~\bibnamefont {Bernard}}, \bibinfo {author} {\bibfnamefont
			{B.}~\bibnamefont {Hambrecht}}, \bibinfo {author} {\bibfnamefont
			{P.}~\bibnamefont {Lugan}}, \bibinfo {author} {\bibfnamefont
			{D.}~\bibnamefont {Cl{\'{e}}ment}}, \bibinfo {author} {\bibfnamefont
			{L.}~\bibnamefont {Sanchez-Palencia}}, \bibinfo {author} {\bibfnamefont
			{P.}~\bibnamefont {Bouyer}}, \ and\ \bibinfo {author} {\bibfnamefont
			{A.}~\bibnamefont {Aspect}},\ }\href {\doibase 10.1038/nature07000}
	{\bibfield  {journal} {\bibinfo  {journal} {Nature}\ }\textbf {\bibinfo
			{volume} {453}},\ \bibinfo {pages} {891} (\bibinfo {year}
		{2008})}\BibitemShut {NoStop}%
	\bibitem [{\citenamefont {Roati}\ \emph {et~al.}(2008)\citenamefont {Roati},
		\citenamefont {D'Errico}, \citenamefont {Fallani}, \citenamefont {Fattori},
		\citenamefont {Fort}, \citenamefont {Zaccanti}, \citenamefont {Modugno},
		\citenamefont {Modugno},\ and\ \citenamefont {Inguscio}}]{Roati2008}%
	\BibitemOpen
	\bibfield  {author} {\bibinfo {author} {\bibfnamefont {G.}~\bibnamefont
			{Roati}}, \bibinfo {author} {\bibfnamefont {C.}~\bibnamefont {D'Errico}},
		\bibinfo {author} {\bibfnamefont {L.}~\bibnamefont {Fallani}}, \bibinfo
		{author} {\bibfnamefont {M.}~\bibnamefont {Fattori}}, \bibinfo {author}
		{\bibfnamefont {C.}~\bibnamefont {Fort}}, \bibinfo {author} {\bibfnamefont
			{M.}~\bibnamefont {Zaccanti}}, \bibinfo {author} {\bibfnamefont
			{G.}~\bibnamefont {Modugno}}, \bibinfo {author} {\bibfnamefont
			{M.}~\bibnamefont {Modugno}}, \ and\ \bibinfo {author} {\bibfnamefont
			{M.}~\bibnamefont {Inguscio}},\ }\href {\doibase 10.1038/nature07071}
	{\bibfield  {journal} {\bibinfo  {journal} {Nature}\ }\textbf {\bibinfo
			{volume} {453}},\ \bibinfo {pages} {895} (\bibinfo {year}
		{2008})}\BibitemShut {NoStop}%
	\bibitem [{\citenamefont {Lahini}\ \emph {et~al.}(2008)\citenamefont {Lahini},
		\citenamefont {Avidan}, \citenamefont {Pozzi}, \citenamefont {Sorel},
		\citenamefont {Morandotti}, \citenamefont {Christodoulides},\ and\
		\citenamefont {Silberberg}}]{Lahini2008}%
	\BibitemOpen
	\bibfield  {author} {\bibinfo {author} {\bibfnamefont {Y.}~\bibnamefont
			{Lahini}}, \bibinfo {author} {\bibfnamefont {A.}~\bibnamefont {Avidan}},
		\bibinfo {author} {\bibfnamefont {F.}~\bibnamefont {Pozzi}}, \bibinfo
		{author} {\bibfnamefont {M.}~\bibnamefont {Sorel}}, \bibinfo {author}
		{\bibfnamefont {R.}~\bibnamefont {Morandotti}}, \bibinfo {author}
		{\bibfnamefont {D.~N.}\ \bibnamefont {Christodoulides}}, \ and\ \bibinfo
		{author} {\bibfnamefont {Y.}~\bibnamefont {Silberberg}},\ }\href {\doibase
		10.1103/PhysRevLett.100.013906} {\bibfield  {journal} {\bibinfo  {journal}
			{Physical Review Letters}\ }\textbf {\bibinfo {volume} {100}},\ \bibinfo
		{pages} {013906} (\bibinfo {year} {2008})}\BibitemShut {NoStop}%
	\bibitem [{\citenamefont {Deissler}\ \emph {et~al.}(2010)\citenamefont
		{Deissler}, \citenamefont {Zaccanti}, \citenamefont {Roati}, \citenamefont
		{D'Errico}, \citenamefont {Fattori}, \citenamefont {Modugno}, \citenamefont
		{Modugno},\ and\ \citenamefont {Inguscio}}]{Deissler2010}%
	\BibitemOpen
	\bibfield  {author} {\bibinfo {author} {\bibfnamefont {B.}~\bibnamefont
			{Deissler}}, \bibinfo {author} {\bibfnamefont {M.}~\bibnamefont {Zaccanti}},
		\bibinfo {author} {\bibfnamefont {G.}~\bibnamefont {Roati}}, \bibinfo
		{author} {\bibfnamefont {C.}~\bibnamefont {D'Errico}}, \bibinfo {author}
		{\bibfnamefont {M.}~\bibnamefont {Fattori}}, \bibinfo {author} {\bibfnamefont
			{M.}~\bibnamefont {Modugno}}, \bibinfo {author} {\bibfnamefont
			{G.}~\bibnamefont {Modugno}}, \ and\ \bibinfo {author} {\bibfnamefont
			{M.}~\bibnamefont {Inguscio}},\ }\href {\doibase 10.1038/nphys1635}
	{\bibfield  {journal} {\bibinfo  {journal} {Nature Physics}\ }\textbf
		{\bibinfo {volume} {6}},\ \bibinfo {pages} {354} (\bibinfo {year}
		{2010})}\BibitemShut {NoStop}%
	\bibitem [{\citenamefont {Gadway}\ \emph {et~al.}(2011)\citenamefont {Gadway},
		\citenamefont {Pertot}, \citenamefont {Reeves}, \citenamefont {Vogt},\ and\
		\citenamefont {Schneble}}]{Gadway2011}%
	\BibitemOpen
	\bibfield  {author} {\bibinfo {author} {\bibfnamefont {B.}~\bibnamefont
			{Gadway}}, \bibinfo {author} {\bibfnamefont {D.}~\bibnamefont {Pertot}},
		\bibinfo {author} {\bibfnamefont {J.}~\bibnamefont {Reeves}}, \bibinfo
		{author} {\bibfnamefont {M.}~\bibnamefont {Vogt}}, \ and\ \bibinfo {author}
		{\bibfnamefont {D.}~\bibnamefont {Schneble}},\ }\href {\doibase
		10.1103/PhysRevLett.107.145306} {\bibfield  {journal} {\bibinfo  {journal}
			{Physical Review Letters}\ }\textbf {\bibinfo {volume} {107}},\ \bibinfo
		{pages} {145306} (\bibinfo {year} {2011})}\BibitemShut {NoStop}%
	\bibitem [{\citenamefont {D'Errico}\ \emph {et~al.}(2014)\citenamefont
		{D'Errico}, \citenamefont {Lucioni}, \citenamefont {Tanzi}, \citenamefont
		{Gori}, \citenamefont {Roux}, \citenamefont {McCulloch}, \citenamefont
		{Giamarchi}, \citenamefont {Inguscio},\ and\ \citenamefont
		{Modugno}}]{DErrico2014}%
	\BibitemOpen
	\bibfield  {author} {\bibinfo {author} {\bibfnamefont {C.}~\bibnamefont
			{D'Errico}}, \bibinfo {author} {\bibfnamefont {E.}~\bibnamefont {Lucioni}},
		\bibinfo {author} {\bibfnamefont {L.}~\bibnamefont {Tanzi}}, \bibinfo
		{author} {\bibfnamefont {L.}~\bibnamefont {Gori}}, \bibinfo {author}
		{\bibfnamefont {G.}~\bibnamefont {Roux}}, \bibinfo {author} {\bibfnamefont
			{I.~P.}\ \bibnamefont {McCulloch}}, \bibinfo {author} {\bibfnamefont
			{T.}~\bibnamefont {Giamarchi}}, \bibinfo {author} {\bibfnamefont
			{M.}~\bibnamefont {Inguscio}}, \ and\ \bibinfo {author} {\bibfnamefont
			{G.}~\bibnamefont {Modugno}},\ }\href {\doibase
		10.1103/PhysRevLett.113.095301} {\bibfield  {journal} {\bibinfo  {journal}
			{Physical Review Letters}\ }\textbf {\bibinfo {volume} {113}},\ \bibinfo
		{pages} {095301} (\bibinfo {year} {2014})}\BibitemShut {NoStop}%
	\bibitem [{\citenamefont {Kondov}\ \emph {et~al.}(2015)\citenamefont {Kondov},
		\citenamefont {McGehee}, \citenamefont {Xu},\ and\ \citenamefont
		{DeMarco}}]{Kondov2015}%
	\BibitemOpen
	\bibfield  {author} {\bibinfo {author} {\bibfnamefont {S.~S.}\ \bibnamefont
			{Kondov}}, \bibinfo {author} {\bibfnamefont {W.~R.}\ \bibnamefont {McGehee}},
		\bibinfo {author} {\bibfnamefont {W.}~\bibnamefont {Xu}}, \ and\ \bibinfo
		{author} {\bibfnamefont {B.}~\bibnamefont {DeMarco}},\ }\href {\doibase
		10.1103/PhysRevLett.114.083002} {\bibfield  {journal} {\bibinfo  {journal}
			{Physical Review Letters}\ }\textbf {\bibinfo {volume} {114}},\ \bibinfo
		{pages} {083002} (\bibinfo {year} {2015})}\BibitemShut {NoStop}%
	\bibitem [{\citenamefont {Gornyi}\ \emph {et~al.}(2005)\citenamefont {Gornyi},
		\citenamefont {Mirlin},\ and\ \citenamefont {Polyakov}}]{Gornyi2005}%
	\BibitemOpen
	\bibfield  {author} {\bibinfo {author} {\bibfnamefont {I.~V.}\ \bibnamefont
			{Gornyi}}, \bibinfo {author} {\bibfnamefont {A.~D.}\ \bibnamefont {Mirlin}},
		\ and\ \bibinfo {author} {\bibfnamefont {D.~G.}\ \bibnamefont {Polyakov}},\
	}\href {\doibase 10.1103/PhysRevLett.95.206603} {\bibfield  {journal}
		{\bibinfo  {journal} {Physical Review Letters}\ }\textbf {\bibinfo {volume}
			{95}},\ \bibinfo {pages} {206603} (\bibinfo {year} {2005})}\BibitemShut
	{NoStop}%
	\bibitem [{\citenamefont {Basko}\ \emph {et~al.}(2006)\citenamefont {Basko},
		\citenamefont {Aleiner},\ and\ \citenamefont {Altshuler}}]{Basko2006}%
	\BibitemOpen
	\bibfield  {author} {\bibinfo {author} {\bibfnamefont {D.}~\bibnamefont
			{Basko}}, \bibinfo {author} {\bibfnamefont {I.}~\bibnamefont {Aleiner}}, \
		and\ \bibinfo {author} {\bibfnamefont {B.}~\bibnamefont {Altshuler}},\ }\href
	{\doibase 10.1016/j.aop.2005.11.014} {\bibfield  {journal} {\bibinfo
			{journal} {Annals of Physics}\ }\textbf {\bibinfo {volume} {321}},\ \bibinfo
		{pages} {1126} (\bibinfo {year} {2006})}\BibitemShut {NoStop}%
	\bibitem [{\citenamefont {Oganesyan}\ and\ \citenamefont
		{Huse}(2007)}]{Oganesyan2007}%
	\BibitemOpen
	\bibfield  {author} {\bibinfo {author} {\bibfnamefont {V.}~\bibnamefont
			{Oganesyan}}\ and\ \bibinfo {author} {\bibfnamefont {D.~A.}\ \bibnamefont
			{Huse}},\ }\href {\doibase 10.1103/PhysRevB.75.155111} {\bibfield  {journal}
		{\bibinfo  {journal} {Physical Review B}\ }\textbf {\bibinfo {volume} {75}},\
		\bibinfo {pages} {155111} (\bibinfo {year} {2007})}\BibitemShut {NoStop}%
	\bibitem [{\citenamefont {Pal}\ and\ \citenamefont {Huse}(2010)}]{Pal2010}%
	\BibitemOpen
	\bibfield  {author} {\bibinfo {author} {\bibfnamefont {A.}~\bibnamefont
			{Pal}}\ and\ \bibinfo {author} {\bibfnamefont {D.~A.}\ \bibnamefont {Huse}},\
	}\href {\doibase 10.1103/PhysRevB.82.174411} {\bibfield  {journal} {\bibinfo
			{journal} {Physical Review B}\ }\textbf {\bibinfo {volume} {82}},\ \bibinfo
		{pages} {174411} (\bibinfo {year} {2010})}\BibitemShut {NoStop}%
	\bibitem [{\citenamefont {Imbrie}(2016)}]{Imbrie2016}%
	\BibitemOpen
	\bibfield  {author} {\bibinfo {author} {\bibfnamefont {J.~Z.}\ \bibnamefont
			{Imbrie}},\ }\href {\doibase 10.1007/s10955-016-1508-x} {\emph {\bibinfo
			{title} {Journal of Statistical Physics}}},\ Vol.\ \bibinfo {volume} {163}\
	(\bibinfo  {publisher} {Springer US},\ \bibinfo {year} {2016})\ pp.\ \bibinfo
	{pages} {998--1048}\BibitemShut {NoStop}%
	\bibitem [{\citenamefont {Abanin}\ \emph {et~al.}(2018)\citenamefont {Abanin},
		\citenamefont {Altman}, \citenamefont {Bloch},\ and\ \citenamefont
		{Serbyn}}]{Abanin2018}%
	\BibitemOpen
	\bibfield  {author} {\bibinfo {author} {\bibfnamefont {D.~A.}\ \bibnamefont
			{Abanin}}, \bibinfo {author} {\bibfnamefont {E.}~\bibnamefont {Altman}},
		\bibinfo {author} {\bibfnamefont {I.}~\bibnamefont {Bloch}}, \ and\ \bibinfo
		{author} {\bibfnamefont {M.}~\bibnamefont {Serbyn}},\ }\href
	{http://arxiv.org/abs/1804.11065} {\  (\bibinfo {year} {2018})},\ \Eprint
	{http://arxiv.org/abs/1804.11065} {arXiv:1804.11065} \BibitemShut {NoStop}%
	\bibitem [{\citenamefont {Schreiber}\ \emph {et~al.}(2015)\citenamefont
		{Schreiber}, \citenamefont {Hodgman}, \citenamefont {Bordia}, \citenamefont
		{Luschen}, \citenamefont {Fischer}, \citenamefont {Vosk}, \citenamefont
		{Altman}, \citenamefont {Schneider},\ and\ \citenamefont
		{Bloch}}]{Schreiber2015}%
	\BibitemOpen
	\bibfield  {author} {\bibinfo {author} {\bibfnamefont {M.}~\bibnamefont
			{Schreiber}}, \bibinfo {author} {\bibfnamefont {S.~S.}\ \bibnamefont
			{Hodgman}}, \bibinfo {author} {\bibfnamefont {P.}~\bibnamefont {Bordia}},
		\bibinfo {author} {\bibfnamefont {H.~P.}\ \bibnamefont {Luschen}}, \bibinfo
		{author} {\bibfnamefont {M.~H.}\ \bibnamefont {Fischer}}, \bibinfo {author}
		{\bibfnamefont {R.}~\bibnamefont {Vosk}}, \bibinfo {author} {\bibfnamefont
			{E.}~\bibnamefont {Altman}}, \bibinfo {author} {\bibfnamefont
			{U.}~\bibnamefont {Schneider}}, \ and\ \bibinfo {author} {\bibfnamefont
			{I.}~\bibnamefont {Bloch}},\ }\href {\doibase 10.1126/science.aaa7432}
	{\bibfield  {journal} {\bibinfo  {journal} {Science}\ }\textbf {\bibinfo
			{volume} {349}},\ \bibinfo {pages} {842} (\bibinfo {year}
		{2015})}\BibitemShut {NoStop}%
	\bibitem [{\citenamefont {Smith}\ \emph {et~al.}(2016)\citenamefont {Smith},
		\citenamefont {Lee}, \citenamefont {Richerme}, \citenamefont {Neyenhuis},
		\citenamefont {Hess}, \citenamefont {Hauke}, \citenamefont {Heyl},
		\citenamefont {Huse},\ and\ \citenamefont {Monroe}}]{Smith2015}%
	\BibitemOpen
	\bibfield  {author} {\bibinfo {author} {\bibfnamefont {J.}~\bibnamefont
			{Smith}}, \bibinfo {author} {\bibfnamefont {A.}~\bibnamefont {Lee}}, \bibinfo
		{author} {\bibfnamefont {P.}~\bibnamefont {Richerme}}, \bibinfo {author}
		{\bibfnamefont {B.}~\bibnamefont {Neyenhuis}}, \bibinfo {author}
		{\bibfnamefont {P.~W.}\ \bibnamefont {Hess}}, \bibinfo {author}
		{\bibfnamefont {P.}~\bibnamefont {Hauke}}, \bibinfo {author} {\bibfnamefont
			{M.}~\bibnamefont {Heyl}}, \bibinfo {author} {\bibfnamefont {D.~A.}\
			\bibnamefont {Huse}}, \ and\ \bibinfo {author} {\bibfnamefont
			{C.}~\bibnamefont {Monroe}},\ }\href {\doibase 10.1038/nphys3783} {\bibfield
		{journal} {\bibinfo  {journal} {Nature Physics}\ }\textbf {\bibinfo {volume}
			{12}},\ \bibinfo {pages} {907} (\bibinfo {year} {2016})}\BibitemShut
	{NoStop}%
	\bibitem [{\citenamefont {Choi}\ \emph {et~al.}(2016)\citenamefont {Choi},
		\citenamefont {Hild}, \citenamefont {Zeiher}, \citenamefont {Schau{\ss}},
		\citenamefont {Rubio-Abadal}, \citenamefont {Yefsah}, \citenamefont
		{Khemani}, \citenamefont {Huse}, \citenamefont {Bloch},\ and\ \citenamefont
		{Gross}}]{Choi2016}%
	\BibitemOpen
	\bibfield  {author} {\bibinfo {author} {\bibfnamefont {J.-y.}\ \bibnamefont
			{Choi}}, \bibinfo {author} {\bibfnamefont {S.}~\bibnamefont {Hild}}, \bibinfo
		{author} {\bibfnamefont {J.}~\bibnamefont {Zeiher}}, \bibinfo {author}
		{\bibfnamefont {P.}~\bibnamefont {Schau{\ss}}}, \bibinfo {author}
		{\bibfnamefont {A.}~\bibnamefont {Rubio-Abadal}}, \bibinfo {author}
		{\bibfnamefont {T.}~\bibnamefont {Yefsah}}, \bibinfo {author} {\bibfnamefont
			{V.}~\bibnamefont {Khemani}}, \bibinfo {author} {\bibfnamefont {D.~A.}\
			\bibnamefont {Huse}}, \bibinfo {author} {\bibfnamefont {I.}~\bibnamefont
			{Bloch}}, \ and\ \bibinfo {author} {\bibfnamefont {C.}~\bibnamefont
			{Gross}},\ }\href {\doibase 10.1126/science.aaf8834} {\bibfield  {journal}
		{\bibinfo  {journal} {Science (New York, N.Y.)}\ }\textbf {\bibinfo {volume}
			{352}},\ \bibinfo {pages} {1547} (\bibinfo {year} {2016})}\BibitemShut
	{NoStop}%
	\bibitem [{\citenamefont {Serbyn}\ \emph
		{et~al.}(2013{\natexlab{a}})\citenamefont {Serbyn}, \citenamefont
		{Papi{\'{c}}},\ and\ \citenamefont {Abanin}}]{Serbyn2013}%
	\BibitemOpen
	\bibfield  {author} {\bibinfo {author} {\bibfnamefont {M.}~\bibnamefont
			{Serbyn}}, \bibinfo {author} {\bibfnamefont {Z.}~\bibnamefont {Papi{\'{c}}}},
		\ and\ \bibinfo {author} {\bibfnamefont {D.~A.}\ \bibnamefont {Abanin}},\
	}\href {\doibase 10.1103/PhysRevLett.110.260601} {\bibfield  {journal}
		{\bibinfo  {journal} {Physical Review Letters}\ }\textbf {\bibinfo {volume}
			{110}},\ \bibinfo {pages} {260601} (\bibinfo {year}
		{2013}{\natexlab{a}})}\BibitemShut {NoStop}%
	\bibitem [{\citenamefont {Serbyn}\ \emph
		{et~al.}(2013{\natexlab{b}})\citenamefont {Serbyn}, \citenamefont
		{Papi{\'{c}}},\ and\ \citenamefont {Abanin}}]{Serbyn2013a}%
	\BibitemOpen
	\bibfield  {author} {\bibinfo {author} {\bibfnamefont {M.}~\bibnamefont
			{Serbyn}}, \bibinfo {author} {\bibfnamefont {Z.}~\bibnamefont {Papi{\'{c}}}},
		\ and\ \bibinfo {author} {\bibfnamefont {D.~A.}\ \bibnamefont {Abanin}},\
	}\href {\doibase 10.1103/PhysRevLett.111.127201} {\bibfield  {journal}
		{\bibinfo  {journal} {Physical Review Letters}\ }\textbf {\bibinfo {volume}
			{111}},\ \bibinfo {pages} {127201} (\bibinfo {year}
		{2013}{\natexlab{b}})}\BibitemShut {NoStop}%
	\bibitem [{\citenamefont {Huse}\ \emph {et~al.}(2014)\citenamefont {Huse},
		\citenamefont {Nandkishore},\ and\ \citenamefont {Oganesyan}}]{Huse2014}%
	\BibitemOpen
	\bibfield  {author} {\bibinfo {author} {\bibfnamefont {D.~A.}\ \bibnamefont
			{Huse}}, \bibinfo {author} {\bibfnamefont {R.}~\bibnamefont {Nandkishore}}, \
		and\ \bibinfo {author} {\bibfnamefont {V.}~\bibnamefont {Oganesyan}},\ }\href
	{\doibase 10.1103/PhysRevB.90.174202} {\bibfield  {journal} {\bibinfo
			{journal} {Physical Review B}\ }\textbf {\bibinfo {volume} {90}},\ \bibinfo
		{pages} {174202} (\bibinfo {year} {2014})}\BibitemShut {NoStop}%
	\bibitem [{\citenamefont {{\v{Z}}nidari{\v{c}}}\ \emph
		{et~al.}(2008)\citenamefont {{\v{Z}}nidari{\v{c}}}, \citenamefont {Prosen},\
		and\ \citenamefont {Prelov{\v{s}}ek}}]{Znidaric2008}%
	\BibitemOpen
	\bibfield  {author} {\bibinfo {author} {\bibfnamefont {M.}~\bibnamefont
			{{\v{Z}}nidari{\v{c}}}}, \bibinfo {author} {\bibfnamefont {T.}~\bibnamefont
			{Prosen}}, \ and\ \bibinfo {author} {\bibfnamefont {P.}~\bibnamefont
			{Prelov{\v{s}}ek}},\ }\href {\doibase 10.1103/PhysRevB.77.064426} {\bibfield
		{journal} {\bibinfo  {journal} {Physical Review B}\ }\textbf {\bibinfo
			{volume} {77}},\ \bibinfo {pages} {064426} (\bibinfo {year}
		{2008})}\BibitemShut {NoStop}%
	\bibitem [{\citenamefont {Bardarson}\ \emph {et~al.}(2012)\citenamefont
		{Bardarson}, \citenamefont {Pollmann},\ and\ \citenamefont
		{Moore}}]{Bardarson2012}%
	\BibitemOpen
	\bibfield  {author} {\bibinfo {author} {\bibfnamefont {J.~H.}\ \bibnamefont
			{Bardarson}}, \bibinfo {author} {\bibfnamefont {F.}~\bibnamefont {Pollmann}},
		\ and\ \bibinfo {author} {\bibfnamefont {J.~E.}\ \bibnamefont {Moore}},\
	}\href {\doibase 10.1103/PhysRevLett.109.017202} {\bibfield  {journal}
		{\bibinfo  {journal} {Physical Review Letters}\ }\textbf {\bibinfo {volume}
			{109}},\ \bibinfo {pages} {017202} (\bibinfo {year} {2012})}\BibitemShut
	{NoStop}%
	\bibitem [{\citenamefont {Aubry}\ and\ \citenamefont
		{Andre}(1980)}]{Aubry1980}%
	\BibitemOpen
	\bibfield  {author} {\bibinfo {author} {\bibfnamefont {S.}~\bibnamefont
			{Aubry}}\ and\ \bibinfo {author} {\bibfnamefont {G.}~\bibnamefont {Andre}},\
	}\href
	{https://www.researchgate.net/publication/265502988{\_}Analyticity{\_}breaking{\_}and{\_}Anderson{\_}localization{\_}in{\_}incommensurate{\_}lattices}
	{\bibfield  {journal} {\bibinfo  {journal} {Ann. Israel Phys. Soc.}\ }\textbf
		{\bibinfo {volume} {3}},\ \bibinfo {pages} {133} (\bibinfo {year}
		{1980})}\BibitemShut {NoStop}%
	\bibitem [{\citenamefont {Iyer}\ \emph {et~al.}(2013)\citenamefont {Iyer},
		\citenamefont {Oganesyan}, \citenamefont {Refael},\ and\ \citenamefont
		{Huse}}]{Iyer2013}%
	\BibitemOpen
	\bibfield  {author} {\bibinfo {author} {\bibfnamefont {S.}~\bibnamefont
			{Iyer}}, \bibinfo {author} {\bibfnamefont {V.}~\bibnamefont {Oganesyan}},
		\bibinfo {author} {\bibfnamefont {G.}~\bibnamefont {Refael}}, \ and\ \bibinfo
		{author} {\bibfnamefont {D.~A.}\ \bibnamefont {Huse}},\ }\href {\doibase
		10.1103/PhysRevB.87.134202} {\bibfield  {journal} {\bibinfo  {journal}
			{Physical Review B}\ }\textbf {\bibinfo {volume} {87}},\ \bibinfo {pages}
		{134202} (\bibinfo {year} {2013})}\BibitemShut {NoStop}%
	\bibitem [{\citenamefont {Bakr}\ \emph {et~al.}(2009)\citenamefont {Bakr},
		\citenamefont {Gillen}, \citenamefont {Peng}, \citenamefont {F{\"{o}}lling},\
		and\ \citenamefont {Greiner}}]{Bakr2009}%
	\BibitemOpen
	\bibfield  {author} {\bibinfo {author} {\bibfnamefont {W.~S.}\ \bibnamefont
			{Bakr}}, \bibinfo {author} {\bibfnamefont {J.~I.}\ \bibnamefont {Gillen}},
		\bibinfo {author} {\bibfnamefont {A.}~\bibnamefont {Peng}}, \bibinfo {author}
		{\bibfnamefont {S.}~\bibnamefont {F{\"{o}}lling}}, \ and\ \bibinfo {author}
		{\bibfnamefont {M.}~\bibnamefont {Greiner}},\ }\href {\doibase
		10.1038/nature08482} {\bibfield  {journal} {\bibinfo  {journal} {Nature}\
		}\textbf {\bibinfo {volume} {462}},\ \bibinfo {pages} {74} (\bibinfo {year}
		{2009})}\BibitemShut {NoStop}%
	\bibitem [{\citenamefont {Islam}\ \emph {et~al.}(2015)\citenamefont {Islam},
		\citenamefont {Ma}, \citenamefont {Preiss}, \citenamefont {{Eric Tai}},
		\citenamefont {Lukin}, \citenamefont {Rispoli},\ and\ \citenamefont
		{Greiner}}]{Islam2015}%
	\BibitemOpen
	\bibfield  {author} {\bibinfo {author} {\bibfnamefont {R.}~\bibnamefont
			{Islam}}, \bibinfo {author} {\bibfnamefont {R.}~\bibnamefont {Ma}}, \bibinfo
		{author} {\bibfnamefont {P.~M.}\ \bibnamefont {Preiss}}, \bibinfo {author}
		{\bibfnamefont {M.}~\bibnamefont {{Eric Tai}}}, \bibinfo {author}
		{\bibfnamefont {A.}~\bibnamefont {Lukin}}, \bibinfo {author} {\bibfnamefont
			{M.}~\bibnamefont {Rispoli}}, \ and\ \bibinfo {author} {\bibfnamefont
			{M.}~\bibnamefont {Greiner}},\ }\href {\doibase 10.1038/nature15750}
	{\bibfield  {journal} {\bibinfo  {journal} {Nature}\ }\textbf {\bibinfo
			{volume} {528}},\ \bibinfo {pages} {77} (\bibinfo {year} {2015})}\BibitemShut
	{NoStop}%
	\bibitem [{\citenamefont {Elben}\ \emph {et~al.}(2018)\citenamefont {Elben},
		\citenamefont {Vermersch}, \citenamefont {Dalmonte}, \citenamefont {Cirac},\
		and\ \citenamefont {Zoller}}]{Elben2017}%
	\BibitemOpen
	\bibfield  {author} {\bibinfo {author} {\bibfnamefont {A.}~\bibnamefont
			{Elben}}, \bibinfo {author} {\bibfnamefont {B.}~\bibnamefont {Vermersch}},
		\bibinfo {author} {\bibfnamefont {M.}~\bibnamefont {Dalmonte}}, \bibinfo
		{author} {\bibfnamefont {J.~I.}\ \bibnamefont {Cirac}}, \ and\ \bibinfo
		{author} {\bibfnamefont {P.}~\bibnamefont {Zoller}},\ }\href {\doibase
		10.1103/PhysRevLett.120.050406} {\bibfield  {journal} {\bibinfo  {journal}
			{Physical Review Letters}\ }\textbf {\bibinfo {volume} {120}},\ \bibinfo
		{pages} {050406} (\bibinfo {year} {2018})}\BibitemShut {NoStop}%
	\bibitem [{\citenamefont {Vosk}\ \emph {et~al.}(2015)\citenamefont {Vosk},
		\citenamefont {Huse},\ and\ \citenamefont {Altman}}]{Vosk2015}%
	\BibitemOpen
	\bibfield  {author} {\bibinfo {author} {\bibfnamefont {R.}~\bibnamefont
			{Vosk}}, \bibinfo {author} {\bibfnamefont {D.~A.}\ \bibnamefont {Huse}}, \
		and\ \bibinfo {author} {\bibfnamefont {E.}~\bibnamefont {Altman}},\ }\href
	{\doibase 10.1103/PhysRevX.5.031032} {\bibfield  {journal} {\bibinfo
			{journal} {Physical Review X}\ }\textbf {\bibinfo {volume} {5}},\ \bibinfo
		{pages} {031032} (\bibinfo {year} {2015})}\BibitemShut {NoStop}%
	\bibitem [{\citenamefont {Potter}\ \emph {et~al.}(2015)\citenamefont {Potter},
		\citenamefont {Vasseur},\ and\ \citenamefont {Parameswaran}}]{Potter2015}%
	\BibitemOpen
	\bibfield  {author} {\bibinfo {author} {\bibfnamefont {A.~C.}\ \bibnamefont
			{Potter}}, \bibinfo {author} {\bibfnamefont {R.}~\bibnamefont {Vasseur}}, \
		and\ \bibinfo {author} {\bibfnamefont {S.~A.}\ \bibnamefont {Parameswaran}},\
	}\href {\doibase 10.1103/PhysRevX.5.031033} {\bibfield  {journal} {\bibinfo
			{journal} {Physical Review X}\ }\textbf {\bibinfo {volume} {5}},\ \bibinfo
		{pages} {031033} (\bibinfo {year} {2015})}\BibitemShut {NoStop}%
	\bibitem [{\citenamefont {Khemani}\ \emph {et~al.}(2017)\citenamefont
		{Khemani}, \citenamefont {Lim}, \citenamefont {Sheng},\ and\ \citenamefont
		{Huse}}]{Khemani2017}%
	\BibitemOpen
	\bibfield  {author} {\bibinfo {author} {\bibfnamefont {V.}~\bibnamefont
			{Khemani}}, \bibinfo {author} {\bibfnamefont {S.~P.}\ \bibnamefont {Lim}},
		\bibinfo {author} {\bibfnamefont {D.~N.}\ \bibnamefont {Sheng}}, \ and\
		\bibinfo {author} {\bibfnamefont {D.~A.}\ \bibnamefont {Huse}},\ }\href
	{\doibase 10.1103/PhysRevX.7.021013} {\bibfield  {journal} {\bibinfo
			{journal} {Physical Review X}\ }\textbf {\bibinfo {volume} {7}},\ \bibinfo
		{pages} {021013} (\bibinfo {year} {2017})}\BibitemShut {NoStop}%
	\bibitem [{\citenamefont {L{\"{u}}schen}\ \emph {et~al.}(2017)\citenamefont
		{L{\"{u}}schen}, \citenamefont {Bordia}, \citenamefont {Scherg},
		\citenamefont {Alet}, \citenamefont {Altman}, \citenamefont {Schneider},\
		and\ \citenamefont {Bloch}}]{Lueschen2017a}%
	\BibitemOpen
	\bibfield  {author} {\bibinfo {author} {\bibfnamefont {H.~P.}\ \bibnamefont
			{L{\"{u}}schen}}, \bibinfo {author} {\bibfnamefont {P.}~\bibnamefont
			{Bordia}}, \bibinfo {author} {\bibfnamefont {S.}~\bibnamefont {Scherg}},
		\bibinfo {author} {\bibfnamefont {F.}~\bibnamefont {Alet}}, \bibinfo {author}
		{\bibfnamefont {E.}~\bibnamefont {Altman}}, \bibinfo {author} {\bibfnamefont
			{U.}~\bibnamefont {Schneider}}, \ and\ \bibinfo {author} {\bibfnamefont
			{I.}~\bibnamefont {Bloch}},\ }\href {\doibase 10.1103/PhysRevLett.119.260401}
	{\bibfield  {journal} {\bibinfo  {journal} {Physical Review Letters}\
		}\textbf {\bibinfo {volume} {119}},\ \bibinfo {pages} {260401} (\bibinfo
		{year} {2017})}\BibitemShut {NoStop}%
	\bibitem [{\citenamefont {Agarwal}\ \emph {et~al.}(2017)\citenamefont
		{Agarwal}, \citenamefont {Altman}, \citenamefont {Demler}, \citenamefont
		{Gopalakrishnan}, \citenamefont {Huse},\ and\ \citenamefont
		{Knap}}]{Agarwal2017}%
	\BibitemOpen
	\bibfield  {author} {\bibinfo {author} {\bibfnamefont {K.}~\bibnamefont
			{Agarwal}}, \bibinfo {author} {\bibfnamefont {E.}~\bibnamefont {Altman}},
		\bibinfo {author} {\bibfnamefont {E.}~\bibnamefont {Demler}}, \bibinfo
		{author} {\bibfnamefont {S.}~\bibnamefont {Gopalakrishnan}}, \bibinfo
		{author} {\bibfnamefont {D.~A.}\ \bibnamefont {Huse}}, \ and\ \bibinfo
		{author} {\bibfnamefont {M.}~\bibnamefont {Knap}},\ }\href {\doibase
		10.1002/andp.201600326} {\bibfield  {journal} {\bibinfo  {journal} {Annalen
				der Physik}\ }\textbf {\bibinfo {volume} {529}},\ \bibinfo {pages} {1600326}
		(\bibinfo {year} {2017})}\BibitemShut {NoStop}%
	\bibitem [{\citenamefont {{De Roeck}}\ and\ \citenamefont
		{Huveneers}(2017)}]{Roeck2017}%
	\BibitemOpen
	\bibfield  {author} {\bibinfo {author} {\bibfnamefont {W.}~\bibnamefont {{De
					Roeck}}}\ and\ \bibinfo {author} {\bibfnamefont {F.}~\bibnamefont
			{Huveneers}},\ }\href {\doibase 10.1103/PhysRevB.95.155129} {\bibfield
		{journal} {\bibinfo  {journal} {Physical Review B}\ }\textbf {\bibinfo
			{volume} {95}},\ \bibinfo {pages} {155129} (\bibinfo {year}
		{2017})}\BibitemShut {NoStop}%
	\bibitem [{\citenamefont {Nandkishore}\ and\ \citenamefont
		{Gopalakrishnan}(2017)}]{Nandkishore2017}%
	\BibitemOpen
	\bibfield  {author} {\bibinfo {author} {\bibfnamefont {R.}~\bibnamefont
			{Nandkishore}}\ and\ \bibinfo {author} {\bibfnamefont {S.}~\bibnamefont
			{Gopalakrishnan}},\ }\href {\doibase 10.1002/andp.201600181} {\bibfield
		{journal} {\bibinfo  {journal} {Annalen der Physik}\ }\textbf {\bibinfo
			{volume} {529}},\ \bibinfo {pages} {1600181} (\bibinfo {year}
		{2017})}\BibitemShut {NoStop}%
	\bibitem [{\citenamefont {Ba{\~{n}}uls}\ \emph {et~al.}(2017)\citenamefont
		{Ba{\~{n}}uls}, \citenamefont {Yao}, \citenamefont {Choi}, \citenamefont
		{Lukin},\ and\ \citenamefont {Cirac}}]{Banuls2017}%
	\BibitemOpen
	\bibfield  {author} {\bibinfo {author} {\bibfnamefont {M.~C.}\ \bibnamefont
			{Ba{\~{n}}uls}}, \bibinfo {author} {\bibfnamefont {N.~Y.}\ \bibnamefont
			{Yao}}, \bibinfo {author} {\bibfnamefont {S.}~\bibnamefont {Choi}}, \bibinfo
		{author} {\bibfnamefont {M.~D.}\ \bibnamefont {Lukin}}, \ and\ \bibinfo
		{author} {\bibfnamefont {J.~I.}\ \bibnamefont {Cirac}},\ }\href {\doibase
		10.1103/PhysRevB.96.174201} {\bibfield  {journal} {\bibinfo  {journal}
			{Physical Review B}\ }\textbf {\bibinfo {volume} {96}},\ \bibinfo {pages}
		{174201} (\bibinfo {year} {2017})}\BibitemShut {NoStop}%
	\bibitem [{\citenamefont {Zupancic}\ \emph {et~al.}(2016)\citenamefont
		{Zupancic}, \citenamefont {Preiss}, \citenamefont {Ma}, \citenamefont
		{Lukin}, \citenamefont {{Eric Tai}}, \citenamefont {Rispoli}, \citenamefont
		{Islam},\ and\ \citenamefont {Greiner}}]{Zupancic2016}%
	\BibitemOpen
	\bibfield  {author} {\bibinfo {author} {\bibfnamefont {P.}~\bibnamefont
			{Zupancic}}, \bibinfo {author} {\bibfnamefont {P.~M.}\ \bibnamefont
			{Preiss}}, \bibinfo {author} {\bibfnamefont {R.}~\bibnamefont {Ma}}, \bibinfo
		{author} {\bibfnamefont {A.}~\bibnamefont {Lukin}}, \bibinfo {author}
		{\bibfnamefont {M.}~\bibnamefont {{Eric Tai}}}, \bibinfo {author}
		{\bibfnamefont {M.}~\bibnamefont {Rispoli}}, \bibinfo {author} {\bibfnamefont
			{R.}~\bibnamefont {Islam}}, \ and\ \bibinfo {author} {\bibfnamefont
			{M.}~\bibnamefont {Greiner}},\ }\href {\doibase 10.1364/OE.24.013881}
	{\bibfield  {journal} {\bibinfo  {journal} {Optics Express}\ }\textbf
		{\bibinfo {volume} {24}},\ \bibinfo {pages} {13881} (\bibinfo {year}
		{2016})}\BibitemShut {NoStop}%
	\bibitem [{\citenamefont {Preiss}\ \emph {et~al.}(2015)\citenamefont {Preiss},
		\citenamefont {Ma}, \citenamefont {Tai}, \citenamefont {Lukin}, \citenamefont
		{Rispoli}, \citenamefont {Zupancic}, \citenamefont {Lahini}, \citenamefont
		{Islam},\ and\ \citenamefont {Greiner}}]{Preiss2015}%
	\BibitemOpen
	\bibfield  {author} {\bibinfo {author} {\bibfnamefont {P.~M.}\ \bibnamefont
			{Preiss}}, \bibinfo {author} {\bibfnamefont {R.}~\bibnamefont {Ma}}, \bibinfo
		{author} {\bibfnamefont {M.~E.}\ \bibnamefont {Tai}}, \bibinfo {author}
		{\bibfnamefont {A.}~\bibnamefont {Lukin}}, \bibinfo {author} {\bibfnamefont
			{M.}~\bibnamefont {Rispoli}}, \bibinfo {author} {\bibfnamefont
			{P.}~\bibnamefont {Zupancic}}, \bibinfo {author} {\bibfnamefont
			{Y.}~\bibnamefont {Lahini}}, \bibinfo {author} {\bibfnamefont
			{R.}~\bibnamefont {Islam}}, \ and\ \bibinfo {author} {\bibfnamefont
			{M.}~\bibnamefont {Greiner}},\ }\href {\doibase 10.1126/science.1260364}
	{\bibfield  {journal} {\bibinfo  {journal} {Science}\ }\textbf {\bibinfo
			{volume} {347}},\ \bibinfo {pages} {1229} (\bibinfo {year}
		{2015})}\BibitemShut {NoStop}%
	\bibitem [{\citenamefont {Hartmann}\ \emph {et~al.}(2004)\citenamefont
		{Hartmann}, \citenamefont {Keck}, \citenamefont {Korsch},\ and\ \citenamefont
		{Mossmann}}]{Hartmann2004}%
	\BibitemOpen
	\bibfield  {author} {\bibinfo {author} {\bibfnamefont {T.}~\bibnamefont
			{Hartmann}}, \bibinfo {author} {\bibfnamefont {F.}~\bibnamefont {Keck}},
		\bibinfo {author} {\bibfnamefont {H.~J.}\ \bibnamefont {Korsch}}, \ and\
		\bibinfo {author} {\bibfnamefont {S.}~\bibnamefont {Mossmann}},\ }\href
	{\doibase 10.1088/1367-2630/6/1/002} {\bibfield  {journal} {\bibinfo
			{journal} {New Journal of Physics}\ }\textbf {\bibinfo {volume} {6}}
		(\bibinfo {year} {2004}),\ 10.1088/1367-2630/6/1/002}\BibitemShut {NoStop}%
	\bibitem [{\citenamefont {Ma}\ \emph {et~al.}(2011)\citenamefont {Ma},
		\citenamefont {Tai}, \citenamefont {Preiss}, \citenamefont {Bakr},
		\citenamefont {Simon},\ and\ \citenamefont {Greiner}}]{Ma2011}%
	\BibitemOpen
	\bibfield  {author} {\bibinfo {author} {\bibfnamefont {R.}~\bibnamefont
			{Ma}}, \bibinfo {author} {\bibfnamefont {M.~E.}\ \bibnamefont {Tai}},
		\bibinfo {author} {\bibfnamefont {P.~M.}\ \bibnamefont {Preiss}}, \bibinfo
		{author} {\bibfnamefont {W.~S.}\ \bibnamefont {Bakr}}, \bibinfo {author}
		{\bibfnamefont {J.}~\bibnamefont {Simon}}, \ and\ \bibinfo {author}
		{\bibfnamefont {M.}~\bibnamefont {Greiner}},\ }\href {\doibase
		10.1103/PhysRevLett.107.095301} {\bibfield  {journal} {\bibinfo  {journal}
			{Physical Review Letters}\ }\textbf {\bibinfo {volume} {107}},\ \bibinfo
		{pages} {095301} (\bibinfo {year} {2011})}\BibitemShut {NoStop}%
	\bibitem [{\citenamefont {Melko}\ \emph {et~al.}(2016)\citenamefont {Melko},
		\citenamefont {Herdman}, \citenamefont {Iouchtchenko}, \citenamefont {Roy},\
		and\ \citenamefont {{Del Maestro}}}]{Melko2016}%
	\BibitemOpen
	\bibfield  {author} {\bibinfo {author} {\bibfnamefont {R.~G.}\ \bibnamefont
			{Melko}}, \bibinfo {author} {\bibfnamefont {C.~M.}\ \bibnamefont {Herdman}},
		\bibinfo {author} {\bibfnamefont {D.}~\bibnamefont {Iouchtchenko}}, \bibinfo
		{author} {\bibfnamefont {P.-N.}\ \bibnamefont {Roy}}, \ and\ \bibinfo
		{author} {\bibfnamefont {A.}~\bibnamefont {{Del Maestro}}},\ }\href {\doibase
		10.1103/PhysRevA.93.042336} {\bibfield  {journal} {\bibinfo  {journal}
			{Physical Review A}\ }\textbf {\bibinfo {volume} {93}},\ \bibinfo {pages}
		{042336} (\bibinfo {year} {2016})}\BibitemShut {NoStop}%
	\bibitem [{\citenamefont {Wiseman}\ and\ \citenamefont
		{Vaccaro}(2003)}]{Wiseman2003}%
	\BibitemOpen
	\bibfield  {author} {\bibinfo {author} {\bibfnamefont {H.~M.}\ \bibnamefont
			{Wiseman}}\ and\ \bibinfo {author} {\bibfnamefont {J.~A.}\ \bibnamefont
			{Vaccaro}},\ }\href {\doibase 10.1103/PhysRevLett.91.097902} {\bibfield
		{journal} {\bibinfo  {journal} {Physical Review Letters}\ }\textbf {\bibinfo
			{volume} {91}},\ \bibinfo {pages} {097902} (\bibinfo {year}
		{2003})}\BibitemShut {NoStop}%
	\bibitem [{\citenamefont {Wolf}\ \emph {et~al.}(2008)\citenamefont {Wolf},
		\citenamefont {Verstraete}, \citenamefont {Hastings},\ and\ \citenamefont
		{Cirac}}]{Wolf2008}%
	\BibitemOpen
	\bibfield  {author} {\bibinfo {author} {\bibfnamefont {M.~M.}\ \bibnamefont
			{Wolf}}, \bibinfo {author} {\bibfnamefont {F.}~\bibnamefont {Verstraete}},
		\bibinfo {author} {\bibfnamefont {M.~B.}\ \bibnamefont {Hastings}}, \ and\
		\bibinfo {author} {\bibfnamefont {J.~I.}\ \bibnamefont {Cirac}},\ }\href
	{\doibase 10.1103/PhysRevLett.100.070502} {\bibfield  {journal} {\bibinfo
			{journal} {Physical Review Letters}\ }\textbf {\bibinfo {volume} {100}},\
		\bibinfo {pages} {070502} (\bibinfo {year} {2008})}\BibitemShut {NoStop}%
\end{thebibliography}
\end{document}